\documentclass[fleqn,usenatbib,useAMS]{mnras}

\usepackage{graphicx}	
\usepackage{amsmath}	
\usepackage{amssymb}	
\usepackage{multicol}        
\usepackage{bm}		
\usepackage{pdflscape}	
\usepackage{threeparttable}
\usepackage{makecell}

\newcommand{\ADtilt}[1]{$\text{AD}_{\text{tilt}}=#1\degr$}

\defcitealias{Stalevski2017}{Paper I}
\defcitealias{Stalevski2019}{Paper II}

\usepackage[T1]{fontenc}
\usepackage{ae,aecompl}

\usepackage{newtxtext,newtxmath}

\title[Dissecting the AGN in Circinus -- III]{Dissecting the active galactic nucleus in Circinus -- III. \textit{VLT/FORS2} polarimetry confirms dusty cone illuminated by a tilted accretion disc}

\author[M. Stalevski et al.]{Marko Stalevski$^{1,2}$%
\thanks{Contact e-mail: \href{mailto:mstalevski@aob.rs}{mstalevski.astro@gmail.com}}%
,  Santiago Gonz\'alez-Gait\'an$^{3}$%
, \DJ or\dj e Savi\'{c}$^{1,4}$%
, Makoto Kishimoto$^{5}$%
, Ana Mour\~ao$^{3}$%
, \newauthor Enrique Lopez-Rodriguez$^{6}$%
and Daniel Asmus$^{7,8}$%
\\
$^{1}$Astronomical Observatory, Volgina 7, 11060 Belgrade, Serbia\\
$^{2}$Sterrenkundig Observatorium, Universiteit Gent, Krijgslaan 281-S9, Gent, 9000, Belgium\\
$^{3}$CENTRA-Centro de Astrof\'{\i}sica e Gravita\c{c}\~ao and Departamento de F\'{\i}sica, Instituto Superior T\'ecnico, \\Universidade de Lisboa, Avenida Rovisco Pais, 1049-001 Lisboa, Portugal\\
$^{4}$Institut d'Astrophysique et de G\'eophysique, Universit\'e de Li\`ege, All\'ee du 6 Ao\^ut 19c, 4000 Li\`ege, Belgium\\
$^{5}$Department of Astrophysics \& Atmospheric Sciences, Kyoto Sangyo University, Kyoto 603-8555, Japan\\
$^{6}$Kavli Institute for Particle Astrophysics \& Cosmology (KIPAC), Stanford University, Stanford, CA 94305, USA\\
$^{7}$Department of Physics \& Astronomy, University of Southampton, Southampton, SO17 1BJ, UK\\
$^{8}$Gymnasium Schwarzenbek, 21493 Schwarzenbek, Germany
}

\date{\today}

\pubyear{2022}

\begin{document}
\label{firstpage}
\pagerange{\pageref{firstpage}--\pageref{lastpage}}
\maketitle

\begin{abstract}
We present polarimetric maps of the Circinus galaxy nucleus in the $BVRI$ bands, obtained with VLT/FORS2. Circinus is the closest Seyfert 2 galaxy and harbours an archetypal obscured active galactic nucleus (AGN). Recent high angular resolution imaging revealed that a major fraction of its mid-infrared (MIR) emission is coming from the polar region. Previously, we demonstrated that these observations are consistent with a model of a compact dusty disc and a hyperboloid shell, resembling a hollow cone on larger scales. Here we focus on the AGN core, up to 40 pc from the central engine, and compare the observations to the radiative transfer models. Polarization maps reveal a conical structure, coinciding with the ionization cone. The wavelength-dependence of the polarization degree indicates that scattering on dust grains is producing polarization. The observed polarization degree ($\approx1-3\%$) is lower than predicted by the models; however, this is only a lower limit, since stellar emission dominates the total flux in the optical. The observed polarization angle ($\approx30\degr$) is reproduced by the model of a dusty disc with a hollow cone that is illuminated by a tilted anisotropic central source. An accretion disc aligned with the ionization cone axis, and alternative dust geometries, such as a paraboloid shell, or a torus enveloped by ambient dust, are inconsistent with the data. We conclude that the optical polarimetric imaging supports earlier evidence for the presence of dust in the polar region, tentatively associated with dusty outflows.
\end{abstract}





\section{Introduction}
\label{sec:intro}

The Circinus galaxy (hereafter Circinus) hosts the nearest obscured active galactic nucleus (AGN) with all the features of an archetypal Seyfert 2 galaxy: strong narrow emission lines in the optical, broad emission lines present in the polarized light \citep{RamosAlmeida2016}, a Compton-thick nucleus \citep{Ricci2015} with a prominent Fe K${\alpha}$ line \citep{Arevalo2014}, and a maser disc seen almost edge-on \citep{Greenhill2003}. A well-defined ionization cone is seen on the western side in the optical; the eastern side of the cone is hidden below the host galaxy disc (inclined by $65\degr$, \citealt{Freeman1977}), but appears to be revealed in the polarized near-infrared (NIR) light \citep{Ruiz2000}. At the distance of $4.2$ Mpc \citep{Freeman1977}, and as the second brightest AGN in the mid-infrared (MIR), Circinus is one of the best targets for high angular resolution observations ($1\arcsec\approx20$ pc). The MID-infrared Interferometric instrument (MIDI) and Multi AperTure mid-Infrared Spectro-Scopic Experiment (MATISSE) at the Very Large Telescope Interferometer (VLTI) resolved the parsec-scale emission of Circinus into two components: one coinciding in size and orientation with the maser discs and another extended in the perpendicular direction, projecting approximately along the ionization cone and the polar axis \citep{Tristram2014, Isbell2022} of the AGN. The polar component was found to be responsible for $\approx80\%$ of the total MIR flux. Furthermore, MIR images we obtained with Very Large Telescope (VLT) spectrometer and imager for the mid-infrared (VISIR) showed that polar emission extends up to $\approx40$ pc on both sides of the nucleus \citep*[][hereafter \citetalias{Stalevski2017}]{Stalevski2017}. 
Until recently, dust in the polar region was neglected in the MIR models, which, according to the simplest form of standard AGN picture, assumed that the dust is confined to the equatorial region in the form of a geometrically and optically thick obscuring structure, dubbed "the torus" \citep{Antonucci1993, Netzer2015}. 
Employing radiative transfer simulations, we put forth a model for the dust structure that can explain the puzzling observations of Circinus: a compact geometrically thin dusty disc and a large scale dusty hollow cone, illuminated by a tilted anisotropic emission of the central engine ("the accretion disc") \citep*[\citetalias{Stalevski2017} and][hereafter \citetalias{Stalevski2019}]{Stalevski2019}). 
The model matches well the morphology, spectral energy distribution (SED) and interferometric data. Moreover, it also naturally fits with geometry constraints imposed by size, shape and orientation of the maser discs and the ionization cone. Furthermore, using this model as a basis, and radiative transfer code \textsc{RefleX} \citep{Paltani-Ricci2017}, \citet{Andonie2022} built a model for the X-ray reprocessed emission which reproduces well the 3-70 keV Chandra and NuSTAR spectra of Circinus, including the complex Fe K$\alpha$ zone and the spectral curvature.

Evidence of dust in the polar region has been found for other AGNs as well, on the scales of a few parsecs \citep{Honig2012,Honig2013,Lopez-Gonzaga2016,Honig2017,Leftley2018} to hundreds of parsecs \citep{Braatz1993,Bock2000,Asmus2016,Asmus2019}. This polar dust emission has been tentatively associated with dusty winds driven away by the radiation pressure \citep{Gallagher2015,Leftley2019,Venanzi2020,Williamson2020} and has led to complementing the usual AGN picture when it comes to the dust structure \citep{Honig2019}. While the recent interferometric observations are the main cause for reinvigorated interest in the topic, the presence of dust in the polar region has been considered for decades. \citet{Netzer-Laor1993} suggested that a number of puzzling features of the broad and narrow line regions can be solved if dust is mixed with the narrow-line-emitting gas. Over the years, a number of theoretical works, from analytical solutions \citep[][]{Konigl-Kartje1994} to hydrodynamics simulations \citep[e.g.][]{Wada2012} investigated the mechanisms of launching and maintaining dusty winds. For a more detailed overview of these works, we point the reader to Section 4.2 of \citetalias{Stalevski2017}.

It is apparent now that for interpretation of the thermal dust emission in AGNs, new models that incorporate the polar wind component are necessary. It is crucial that this new generation of models has a well-established basis. In that regard, the AGN in Circinus fulfils all the requirements to serve as a prototype. 
Thus, to further probe and establish the prototype of the dusty wind model, we followed up this source with additional observations using several instruments of the VLT in the IR (in the sparse aperture masking and coronagraphy modes) and optical (integral field spectra and imaging polarimetry). Here we present polarimetric observations of Circinus using the FOcal Reducer/low dispersion Spectrograph 2 (FORS2), a multi mode (imaging, polarimetry, long slit and multi-object spectroscopy) optical instrument mounted on the VLT UT1 Cassegrain focus \citep{Appenzeller1998}. Observations and data reduction are presented in Sec.~\ref{sec:obs}. In previous works \citepalias{Stalevski2017,Stalevski2019}, aside from the model described above, we tested the alternative geometries which, in principle, could explain the emission in the polar region: one where the dust is distributed within the paraboloid shell, and another in which the polar elongation is produced by illumination of the host galaxy ambient dust. We demonstrated that these alternatives are not consistent with the data. However, we consider it is useful to explore the polarization signatures of different geometries and to test if the polarimetric observations alone could distinguish between them. Thus, in Sec.~\ref{sec:mod} we present a limited parameter study of the different radiative transfer models. Then we compare the models to the data (Sec.~\ref{sec:res}) and summarize the results (Sec.~\ref{sec:sum}).
In the present work, our focus is on the AGN core and ionization cone, roughly the central $\approx100\times100$ ${\rm pc}^2$ region. A detailed analysis of the host galaxy polarization and its dust properties will be presented in a follow-up publication.

\section{Observations}
\label{sec:obs}

In this section we present the VLT/FORS2 linearly polarized imaging data obtained as part of the programmes 0101.B-0647 and 0103.B-0517, as well as the reduction techniques and the instrumental and foreground/background corrections we performed. 

\subsection{Observation strategy}

We observed the Circinus galaxy (R.A.$=14^h13^m9.9^s$, Decl.$=-65^{\circ}20'21''$, equinox  J2000) with the FORS2 instrument in the imaging polarimetric mode (IPOL) in the four optical filters $B$, $V$, $R$, and $I$. These observations are achieved by introducing a Wollaston prism that divides the beam into two beams with perpendicular polarizations. A mask with nine 22\arcsec strips ensures that light rays that are perpendicularly polarized  do not overlap on the CCD.  A rotatable half-wave plate (HWP)  placed  at different angles allows for the measurements of the light polarization. We used four HWP position angles ($0^{\circ}$, $22.5^{\circ}$, $45^{\circ}$ and $67.5^{\circ}$) for each set observations.

In $B$-band, we took 20 exposures of 30 seconds (for each of the four HWP angles) with the Circinus galaxy always at the same central position in the CCD. In the other three bands, $VRI$, we took 20, 30, 70 exposures of 80, 30, 20 seconds (for every set of four HWP angles) dithered with 18, 27, 26 different offsets, respectively (see Table~\ref{tab:obs}). The different offsets in declination (up to 3') were taken in order to dither the galaxy and the AGN to different $y$-positions on the CCD (always at the same fixed position angle), which in principle should reduce any spurious spatial effects of the instrument (see Sec.~\ref{subsec:inst}). Observing conditions prevented us from acquiring different offsets for the $B$-band. The mask was placed so that the centre part of the galaxy, where the AGN is located, was always visible within a single strip, regardless of the dithering. Even though the field-of-view of FORS2 ($6.8'\times6.8'$) is in principle large enough to cover the entire Circinus galaxy, some non-central parts were missed due to the presence of the mask. All images taken with a given filter at the same dithering position (and HWP angle) were stacked together following a procedure described bellow. The image pixel scale was 0.252\arcsec/pix (taking already into account the 2$\times$2 binning) which at the distance of Circinus (4.2 Mpc) translates into 19.39 pc for 1\arcsec (or 4.89 pc/pix).

\begin{table*}
\centering
\caption{Summary of observations}
\begin{threeparttable}
\begin{tabular}{c|cccccc}
\hline
\hline
Filter & Exposure time (s) & Nr of exposures & Nr of offsets & Observing nights & FWHM (pix) & FWHM (\arcsec)\tnote{\dag}\\
\hline
$B$ & 30 & 20 & 1 & 2018-05-17 & 2.72 & 0.68\\
$V$ & 80 & 20 & 18 & 2018-06-15, 2018-07-09 & $5.30\pm0.83$ & $1.34\pm0.21$\\
$R$ & 30 & 30 & 27 & \makecell{2018-07-09, 2019-04-07,2019-04-10,\\ 2019-06-29, 2020-02-24, 2020-02-28} & $3.58\pm0.18$ & $0.9\pm0.04$\\
$I$ & 20 & 70 & 26 & 2018-07-13, 2019-04-11 & $3.44\pm0.23$ & $0.87\pm0.06$\\
\end{tabular}
\begin{tablenotes}
\item \dag Median and deviation of all offsets' full width half maximum (FWHM) measured from the brightest stars in the field.
\end{tablenotes}
\end{threeparttable}
\label{tab:obs}
\end{table*}

\subsection{Reduction pipeline}
\label{sec:reduction}

We closely followed the pipeline developed in \citet{Gonzalez-Gaitan20} for imaging polarimetry of extended sources with VLT/FORS2. First, all images were bias-corrected with a master-bias obtained from a stack of all bias frames taken during the night. Cosmic ray removal was done using the {\sc python} package {\sc lacosmic}\footnote{\url{http://www.astro.yale.edu/dokkum/lacosmic/}} \citep{vanDokkum01}. 

The next step aims at measuring separately the fluxes of the two perpendicularly polarized components of the light, called the ordinary and extraordinary beam. To achieve this we separated each image, taken at a given position angle of the HWP, into two images, one corresponding to the ordinary and the other to the extraordinary beam. After that we match their positions, and merged the images of the CCD1 and CCD2 ending thus with two entire new images, one corresponding to the ordinary and another to the  extraordinary beams. We repeated this same approach for images obtained at each of the four HWP angles.

From these eight images we subsequently calculated the Stokes parameters at each pixel position of the masked field-of-view. To do this, one can use the double difference method or the double ratio method \citep[e.g.][]{Patat06,Bagnulo09}. We employed the former but we note that using the double ratio method resulted in virtually no difference in the obtained Stokes parameters. A review of these calculations can be found in App.~\ref{app:stokes}.

The measured normalised Stokes parameters obtained in this way at each position $x,y$ of the field, $q_{\rm obs} (x,y) =Q_{\rm obs}/I_{\rm obs}$ and $u_{\rm obs}(x,y)=U_{\rm obs}/I_{\rm obs}$, where $I_{\rm obs}$ is the intensity,  contain several contaminating factors that need to be corrected for, in order to obtain the real polarization of the Circinus AGN. These corrections, going backwards of the beam arrival, included: the instrumental chromatism of the HWP leading to a wavelength-dependent rotation of the polarization vector (see FORS2 manual\footnote{\url{http://www.eso.org/sci/facilities/paranal/instruments/fors/doc/VLT-MAN-ESO-13100-1543_P01.1.pdf}}), the spatial instrumental polarization induced at the collimator lens \citep{Gonzalez-Gaitan20}, the insterstellar polarization from the Milky Way (MW), the interstellar polarization from the Circinus host galaxy, and a background sky polarization --- which may have various different origins. After the chromatism correction, the latter corrections on the observed Stokes parameters for each wavelength can thus be approximated as:

\begin{equation}\label{eq:stokescorr}
\begin{aligned}
q_{\rm obs} (x, y) = q_{\rm inst}(x,y) + q_{\rm ISP_{\rm MW}} + q_{\rm ISP_{\rm host}} + \frac{Q_{\rm AGN} + Q_{\rm stellar} + Q_{\rm sky}}{I_{\rm AGN} + I_{\rm stellar} + I_{\rm sky}}\\ 
u_{\rm obs} (x, y) = u_{\rm inst}(x,y) + u_{\rm ISP_{\rm MW}} + u_{\rm ISP_{\rm host}} + \frac{U_{\rm AGN} + U_{\rm stellar} + U_{\rm sky}}{I_{\rm AGN} + I_{\rm stellar} + I_{\rm sky}},
\end{aligned}
\end{equation}

\noindent where $q_i = Q_i/I_i$, $u_i = U_i/I_i$ are the normalized Stokes parameters that will be discussed in the next sections. The first three terms in the equation represent processes that may affect the state of the polarization of the traversing photons and are named instrumental (inst), Milky Way (MW) and Circinus host galaxy (host).  The latter term in the equation accounts for the different sources of photons, namely the AGN, the stellar and the sky. Since the stellar light of the host is intrinsically unpolarized, $Q_{\rm stellar}$ can be considered to be zero. Using equation ~(\ref{eq:stokescorr}) it is possible to obtain the Stokes parameters of the AGN: $q_{\rm AGN} = Q_{\rm AGN}/I_{\rm AGN}$ and $u_{\rm AGN} = U_{\rm AGN}/I_{\rm AGN}$  and the actual AGN polarization. However it is necessary to reliably determine $I_{\rm sky}$ and $I_{\rm stellar}$; otherwise, only a lower limit can be inferred (Sec.~\ref{sec:modVsobs}).

When several ditherings were taken (for $VRI$ bands), we first calculated the Stokes parameters individually and made the appropriate corrections. In particular, the instrumental spatial correction changed with the field position of the AGN in each dithering. Only after applying all corrections, we aligned the various offsets (with the {\sc astroalign} module -- \citealt{astroalign20}), and stacked them by taking the median of all matched $q_{\rm AGN}^i(x,y)$ and $u_{\rm AGN}^i(x,y)$.   

Lastly, we performed the polarization bias correction to avoid positive biases of the polarization degree at low signal-to-noise \citep{Serkowski58}. Here, we applied the correction suggested by \citet{Plaszczynski14}.

\subsubsection{Instrumental field correction}\label{subsec:inst}

\begin{figure*}
 \includegraphics[trim={0cm 0 3.5cm 0 },clip,width=0.4885\textwidth]{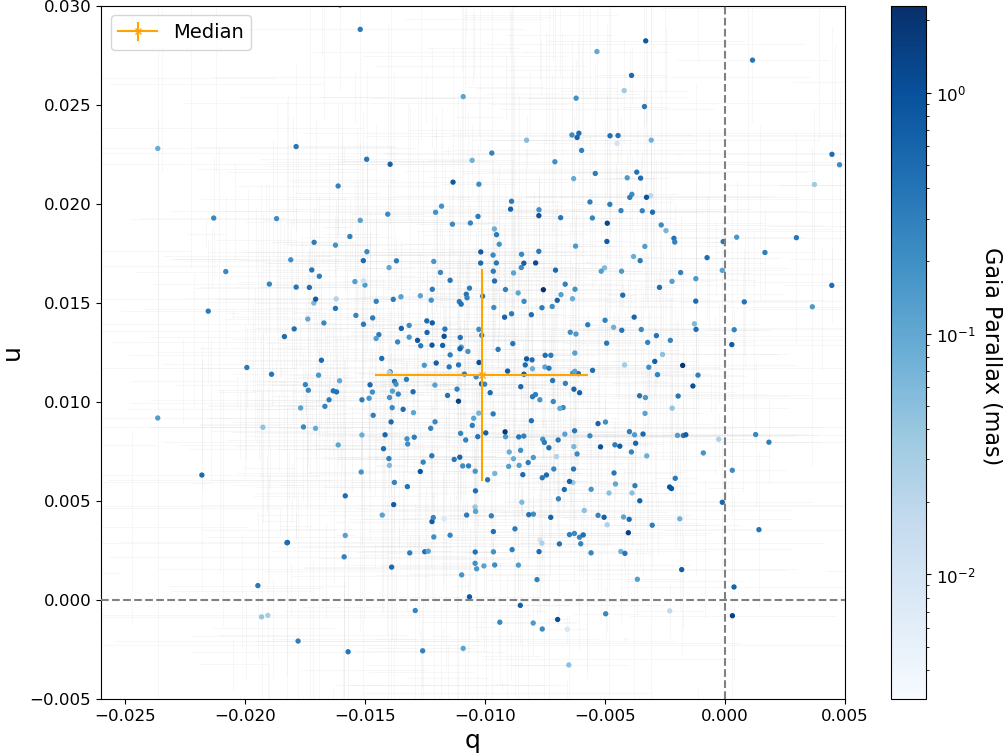}
 \includegraphics[trim={2.5cm 0 0 0},clip,width=0.5115\textwidth]{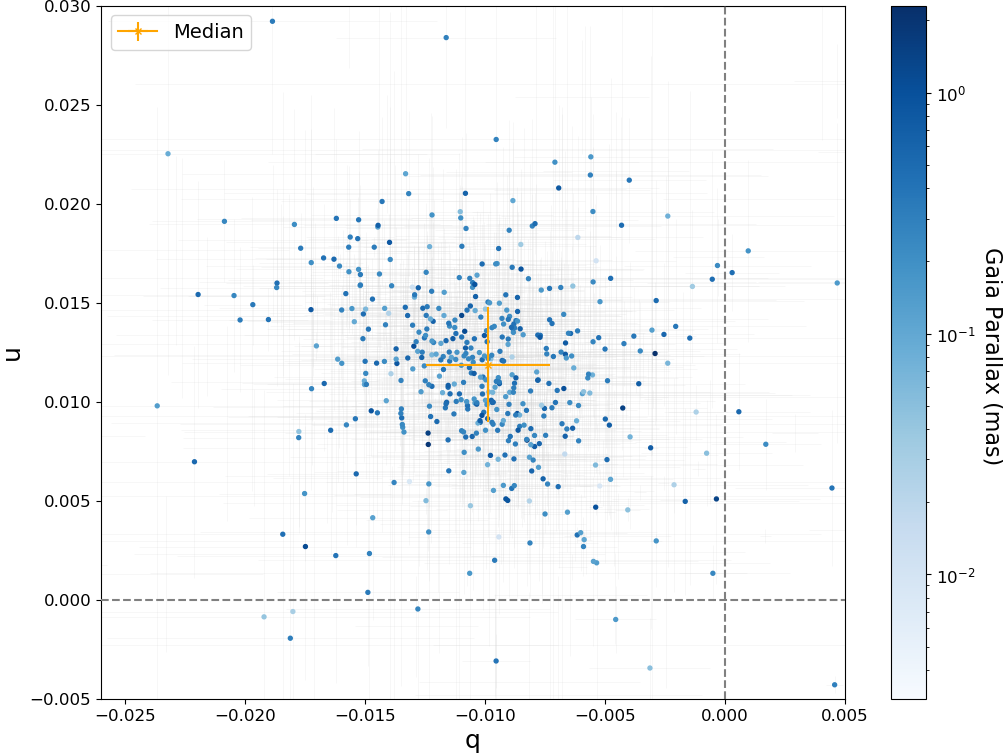}
\caption{Normalised $q$-$u$ plane of the field stars matched with the Gaia catalogue in one of the $R$-band offsets of Circinus field of view without (\emph{left}) and with (\emph{right}) instrumental field correction (see Sec.~\ref{subsec:inst}). The Stokes parameters were obtained from intensities calculated with aperture photometry. The colour bar indicates the Gaia parallax (in milliarcseconds) and the orange point is the median of all stars used to calculate the MW ISP (see Sec.~\ref{subsec:MW-ISP}). The dashed lines indicate where $q=0$ and $u=0$.}
 \label{fig:qu-plane}
\end{figure*}

There is a known spurious instrumental polarization in FORS2 that possibly originates in the collimator lens. This spurious instrumental polarization follows a wavelength-dependent radial pattern reaching up to 1.4\% at the edges and it can be corrected for with analytical functions or non-parametric maps as shown in \citep{Gonzalez-Gaitan20}. We followed the latter approach and noted a clear change  in the $q$-$u$ plane of the field stars (see Fig.~\ref{fig:qu-plane}). 
A similar spurious instrumental polarization had also been found at the FORS1, a detector similar to FORS2 that was earlier installed at the VLT \citep{Patat06}. 

\subsubsection{Interstellar Milky Way polarization}
\label{subsec:MW-ISP}

The Circinus field of view for a given offset contains $\approx800$ stars that allowed us to do a thorough study of the interstellar polarization (ISP) from the MW. By doing photometry on all field stars for both ordinary and extraordinary beams in all four HWP angles, we calculated the Stokes parameters for every star in the field. The photometry was done both with aperture photometry and Point Spread Function (PSF) photometry using the {\sc photutils} package \citep{photutils20}. We corrected for the sky (with iterative source-masked sigma-clipped averages of 50 pixel boxes) prior to doing photometry. For aperture photometry we used a radius equal to the full width half maximum, which is close to the optimal aperture radius that ensures the highest signal-to-noise ratio \citep[e.g.][]{Mighell99}. Since the $q$ and $u$ parameters depend on the ratios of the fluxes, it is unimportant that the flux is not aperture-corrected, as long as the same radius is used for both, ordinary and extraordinary beams. Regarding PSF photometry, we used a simple Gaussian kernel fitted to the 20 brightest stars in the field. The resulting $q$-$u$ plane for a single offset in $R$-band is shown in Fig.~\ref{fig:qu-plane}. 

\begin{figure}
 \includegraphics[width=0.5\textwidth]{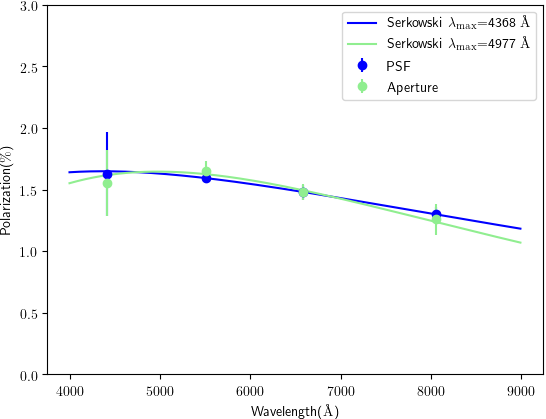}
 
\caption{Interstellar polarization degree percentage from the Milky Way toward Circinus for $BVRI$ obtained from the median of the field stars in several different ditherings with two different photometry methods: aperture and PSF. Corresponding Serkowski fits show that the maximum wavelength of polarization agrees with typical MW values.}
 \label{fig:MW-ISP}
\end{figure}

We matched the stars with the third Gaia release \citep{Gaia16,Gaia20} and we did not find any trend with parallax nor stellar flux in the $q$-$u$ plane. We therefore used the median of all found stars as an estimate of the MW ISP for a given offset. Since we have multiple offsets for three filters, we did this same study for each of those ditherings, correcting for chromatism and instrumental polarization for each (see previous Section). Finally, we took the median value of the Stokes parameters $q$ and $u$ of all offsets to find an optimal value at each sky position and for each filter. The respective uncertainties come from the median absolute deviation of all offsets. For the $B$-band, which has only one offset, the ISP and uncertainties were taken from the median and median absolute deviation of all stars in the field. The final ISP of the MW is shown in Table~\ref{tab:MW-ISP} and in Fig.~\ref{fig:MW-ISP} for both PSF and aperture photometry, which agree very well with each other. We also performed a phenomenological Serkowski fit \citep{Serkowski75} obtaining a wavelength of maximum polarization ($4500-5000$\AA) that agrees well with typical MW values and point towards standard reddening law with $R_V\sim 3.1$ \citep[e.g.][]{Fitzpatrick07,Cikota17}.

Additionally, to confirm the obtained values, we explored a completely independent method based on the dust emission maps from Planck \citep{Planck15,Planck20}. Similarly to \citet{Skalidis19}, we took advantage of the tight correlation between the polarized flux, $P_{\mathrm{submm}}$, in the submillimetre, and the degree of optical polarization of stars in the MW, $p_V$:

\begin{equation}
R_{P/p} = \frac{P_{\mathrm{submm}}}{p_V} = [5.4 \pm 0.2 \mathrm{(stat)} \pm 0.3 \mathrm{(syst)}] \,\mathrm{MJy/sr},    
\end{equation}
or alternatively:

\begin{equation}
R_{S/V} = \frac{P_{\mathrm{submm}}/I_{\mathrm{submm}}}{p_V/\tau_V} = 4.2 \pm 0.2 \mathrm{(stat)} \pm 0.3 \mathrm{(syst)},    
\end{equation}
where $\tau_V$ is the optical depth of the stars proportional to their colour excess, $E(B-V)$, and the MW reddening law $R_V=3.1$: $\tau_V = A_V/1.086 = E(B-V)\cdot R_V/1.086$. For the second relation we used the colour excess from the maps of \citet{Schlafly11} at the centre position of Circinus and its associated error from the deviation of the full field-of-view, $E(B-V)=1.46\pm0.14$. The Planck Stokes maps at the desired coordinate are (MJy/sr): $Q_{\mathrm{submm}}=5.87\pm0.07\times10^{-2}$,  $U_{\mathrm{submm}}=4.62\pm0.10\times10^{-2}$ and $I_{\mathrm{submm}}=4.048\pm0.001$. We obtained an optical degree of polarization of $p_V(\%)=1.33\pm 0.07$ using the first method and $p_V(\%)=1.83\pm 0.23$ for the second. These take into account uncertainties in the Stokes parameters, in the ratios $R$, and in the colour excess. The angle of polarization -- shifted 90$^{\circ}$ with respect to the emission -- is: $\chi{_V}(^{\circ})=71.0 \pm 1.6$ for the first case and $\chi{_V}(^{\circ})=70.9 \pm 0.4$ for the second. So, although the Planck maps have a lower resolution ($\sim10'$ at lower Galactic latitudes) and the results change somewhat with the ratio method used, this result is remarkably consistent with the $V$-band values we obtained directly from stars in the field (see Table~\ref{tab:MW-ISP}), confirming the accuracy of our results.
We also note that there is a discrepancy of around $-4.7^{\circ}\pm2.0^{\circ}$ which actually agrees with the systematic $-3.1^{\circ}$ offset found in \citet{Planck20}.

\begin{table}
\centering
\caption{Milky Way interstellar polarization towards Circinus}
\begin{tabular}{c|cc|cc}
\hline
\hline
Filter & \multicolumn{2}{c}{Aperture} & \multicolumn{2}{c}{PSF} \\
 & $p$(\%) & $\chi$(deg) & $p$(\%) & $\chi$(deg) \\
\hline
$B$ & 1.55$\pm$0.27 & 69.0$\pm$4.9 & 1.63$\pm$0.34 & 67.0$\pm$5.9\\
$V$ & 1.65$\pm$0.09 & 66.2$\pm$1.6 & 1.59$\pm$0.04 & 66.2$\pm$0.7\\
$R$ & 1.48$\pm$0.07 & 63.0$\pm$1.4 & 1.48$\pm$0.06 & 63.6$\pm$1.2\\
$I$ & 1.26$\pm$0.13 & 61.4$\pm$4.2 & 1.30$\pm$0.06 & 62.4$\pm$1.2\\
\end{tabular}
\label{tab:MW-ISP}
\end{table}

\subsubsection{Sky background correction}\label{subsec:sky}

The sky background in the Circinus field of view presents a polarization that is not constant and changes according to the date, filter and offset of the observations. Although all our data were taken in dark, clear nights and pointing far away from the ecliptic, avoiding thus contamination from moonlit and zodiacal sky polarization, we found a sky polarization degree in a range from 0 to 5\% of unknown origin (see Table~\ref{tab:sky}).

To obtain these sky polarization values, we masked out all stars and the entire Circinus galaxy by removing an isophotal ellipse with a semi-major axis of $\sim$670 pixels (i.e. $\sim84.8\arcsec$)\footnote{We used the {\sc isophote} module of {\sc photutils} \citep{photutils20}}. We then binned the remaining pixel values of ordinary and extraordinary beams in boxes of 30$\times$30 pixels by taking the 2$\sigma$-clipped median and standard deviation of each. From these binned fluxes, we calculated the normalised Stokes parameters $q$ and $u$ (see App.~\ref{app:stokes}) and corrected them for chromatism and instrumental polarization (see Sec.~\ref{subsec:inst}). 

We find that the sky polarization changes substantially with the filter and offset (see columns 3 and 5 of Table~\ref{tab:sky}) and we therefore corrected each $q$ and $u$ maps from a given offset individually using Eq.~\ref{eq:stokescorr}. It also changes throughout the field but much less (see columns 4 and 6). Since the exact spatial variations of the background are difficult to estimate, particularly at the AGN position that has been masked out, we assumed a constant background value by taking the median of the entire masked image for a given offset. The uncertainties in this sky correction are greatly reduced given the low sky intensity compared to the AGN (see column 1 of Table~\ref{tab:sky}). The unnormalised Stokes parameters ($Q_{\rm sky}$, $U_{\rm sky}$) that enter the sky correction (Eq.~\ref{eq:stokescorr}) only contribute up to 4\% of the Circinus Stokes parameters.

\begin{table}
\centering
\caption{Sky polarization towards Circinus}
\label{tab:sky}
\scalebox{0.83}{
\begin{threeparttable}
\begin{tabular}{c|c|cc|cc}
\hline
\hline
Filter & $I_{\rm sky}/I_{\rm AGN}$\tnote{$\diamond$} & $p_{\rm sky}$(\%)\tnote{$\circ$} & $\delta p_{\rm sky}$(\%)\tnote{\dag} & $\chi_{\rm sky}$(deg)\tnote{$\square$} & $\delta\chi_{\rm sky}$(deg)\tnote{$\triangle$} \\
\hline
$B$\tnote{$\ast$} & 0.038 & 2.67 & 0.46 & 151.2 & 4.0\\
$V$ & $0.020 \pm 0.002$ & $3.86\pm0.79$ & $0.42\pm0.07$ & 167.4$\pm$7.8 & 4.1$\pm$0.7\\
$R$ & $0.013 \pm 0.003$ & $1.90\pm1.38$ & $0.41\pm0.06$ & 90.5$\pm$71.9 & 21.3$\pm$23.6\\
$I$ & $0.020 \pm 0.002$& $0.35\pm0.59$ & $0.49\pm0.12$ & $116.9 \pm 52.5$ & $66.6\pm41.5$ \\
\end{tabular}
\begin{tablenotes}
\item $\diamond$ Average and standard deviation for all offsets of the ratio of sky intensity to central pixel AGN intensity
\item $\circ$ Average and standard deviation for all offsets of the mean field sky polarization degree
\item \dag Average and standard deviation for all offsets of the dispersion in the field sky polarization
\item $\square$ Average and standard deviation for all offsets of the mean field sky polarization angle
\item $\triangle$ Average and standard deviation for all offsets of the dispersion of the field sky polarization angle. 
\item $\ast$ Note that $B$-band has only one offset.
\end{tablenotes}
\end{threeparttable}
}
\end{table}

\subsubsection{Interstellar polarization of the host galaxy}

The polarization due to dichroic absorption by the aligned dust grains changes across the host galaxy depending on the magnetic and radiation fields responsible for the alignment. Therefore, it is not as straightforward to correct for as for the uniform MW ISP.
However, as discussed in the following section, there are indications that the region of our interest -- AGN ionization cone --  is dominated by polarization due to scattering on dust grains. Thus, in the present work, we did not attempt to correct for the host ISP. We will present several approaches in the host ISP treatment in the forthcoming companion paper.

\subsubsection{Polarization in the nuclear region of Circinus}
\label{sec:obs-maps}

As detailed in the Sec.~\ref{sec:reduction}, after correcting every pixel for chromatism, instrumental polarization, MW ISP polarization and sky polarization, as well as applying the polarization bias correction, we obtained the final Stokes parameters.
It is important to emphasize that, while some of the corrections (e.g. MW ISP) are on the level of inferred Circinus AGN polarization, their associated uncertainties are in all cases one or two magnitudes lower (see tables \ref{tab:MW-ISP}, \ref{tab:sky}, \ref{tab:con}).

\begin{figure*}
 \includegraphics[width=\textwidth]{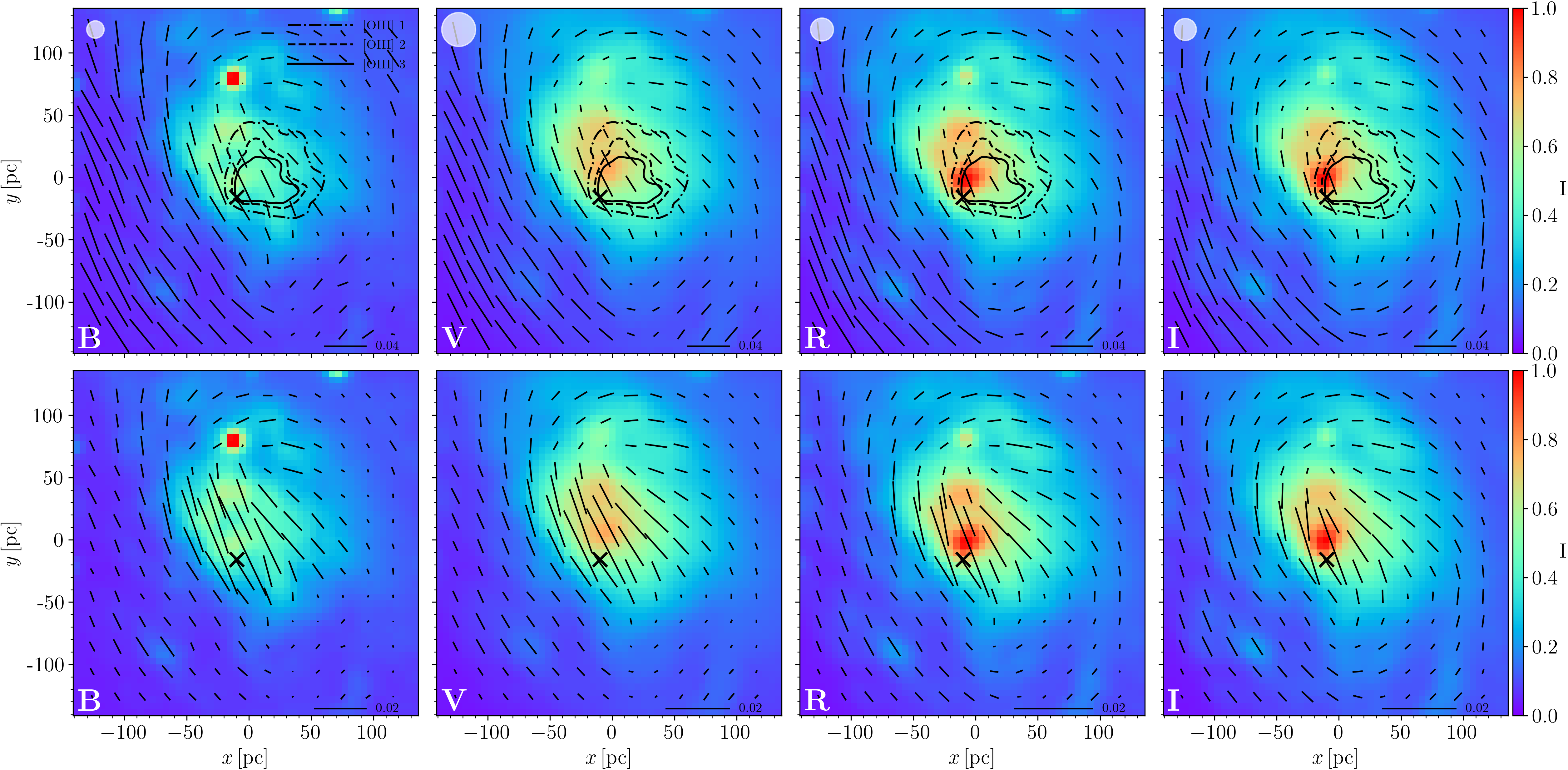}
 \caption{Normalized flux of the central $280\times280$ ${\rm pc}^2$ in the $BVRI$ filters (color maps from left to right). Overplotted are the polarization vectors in $4\times4$ pixel bins with their length corresponding to the polarization degree (top row) and polarized flux (bottom row), see text for details. Also shown are the [\ion{O}{iii}] contours from VLT/MUSE data (Kakkad et al, subm.). X marks the position of the AGN nucleus. White circle in the top left corner corresponds to the median FWHM of the PSF in each band. In all the maps, Norht is up, East is left.}
 \label{fig:obsPolVec}
\end{figure*}

\begin{figure*}
 \includegraphics[width=0.49\textwidth]{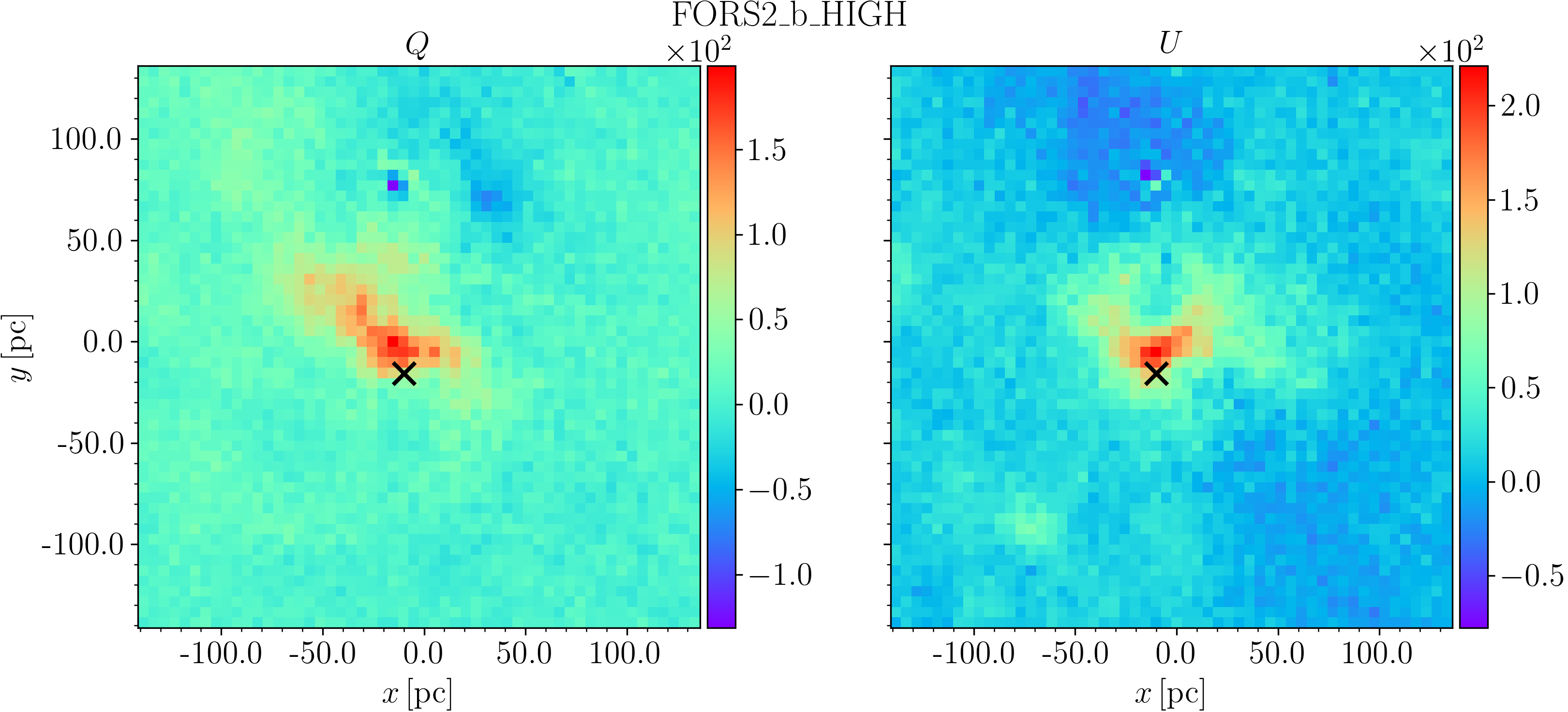}
 \includegraphics[width=0.49\textwidth]{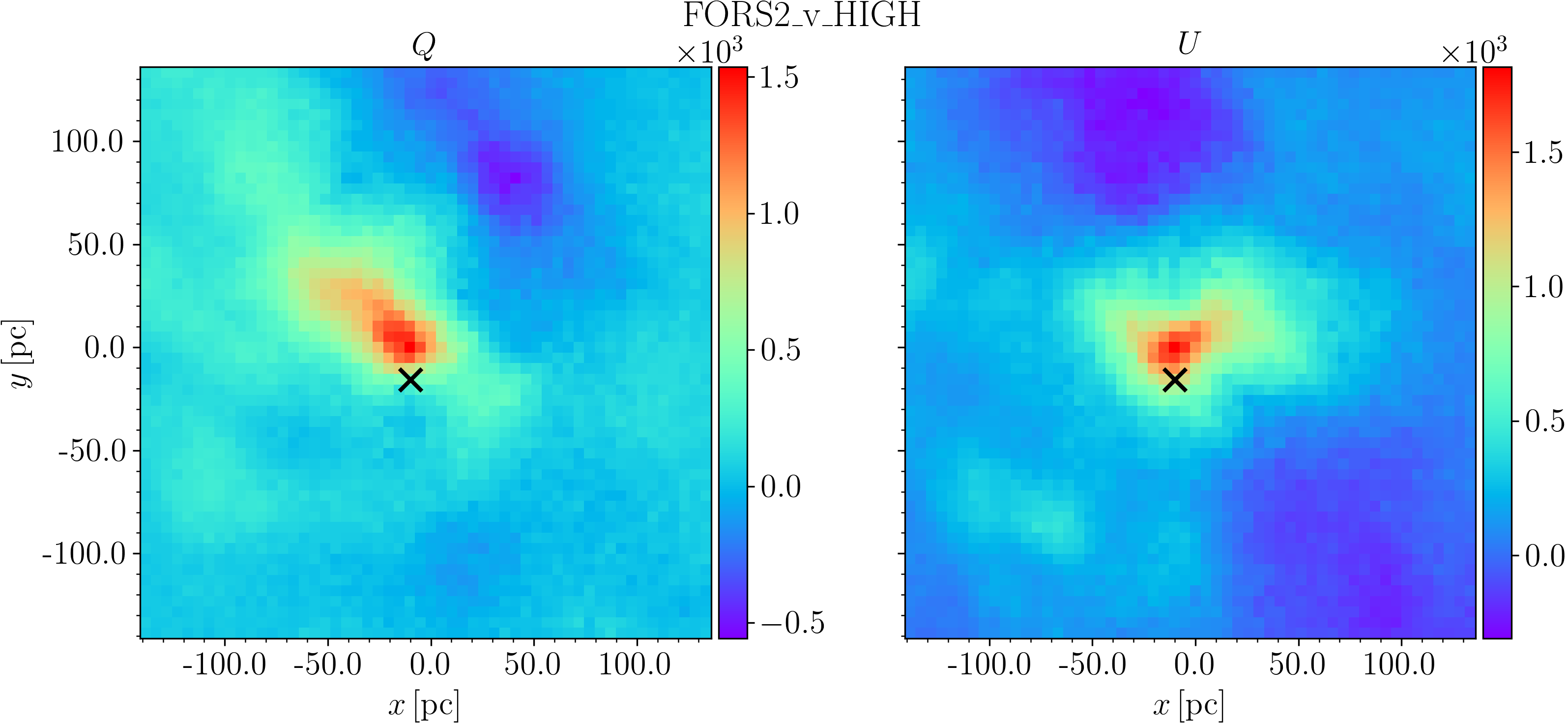}\\
 \includegraphics[width=0.49\textwidth]{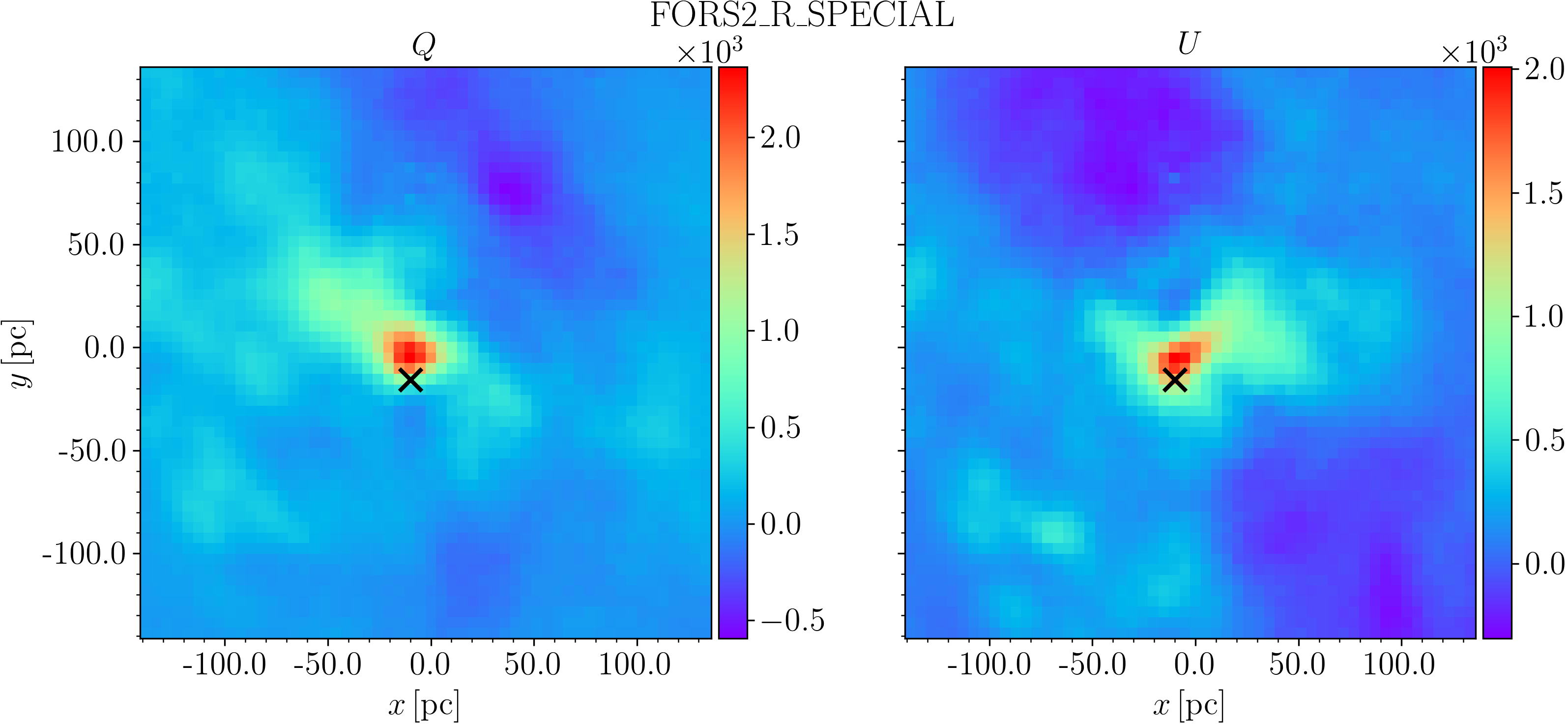}
 \includegraphics[width=0.49\textwidth]{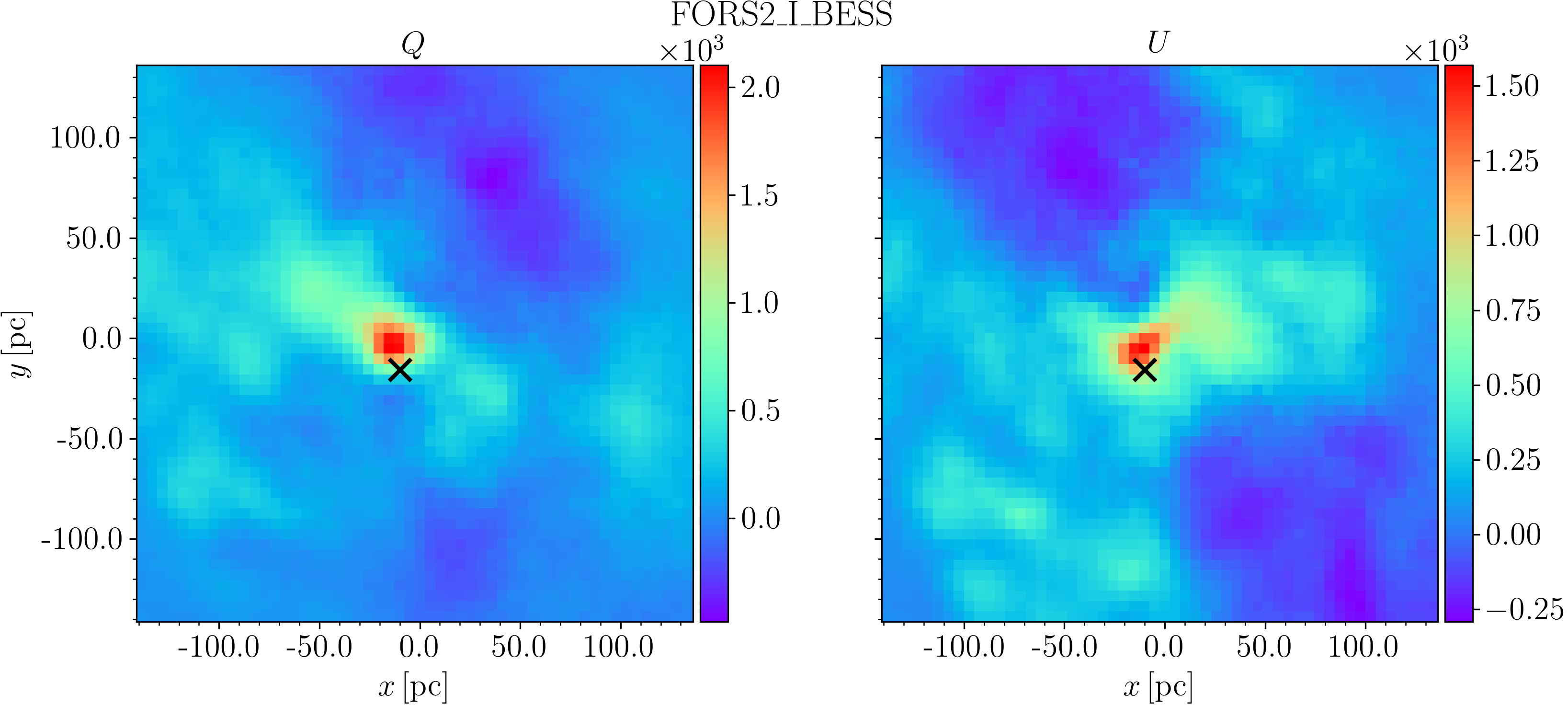}
 \caption{Stokes $Q$ and $U$ parameters maps of the central $280\times280$ ${\rm pc}^2$ in the $BVRI$ bands. X marks the position of the AGN nucleus.}
 \label{fig:obsQU}
\end{figure*}

\begin{figure*}
 \includegraphics[width=0.49\textwidth]{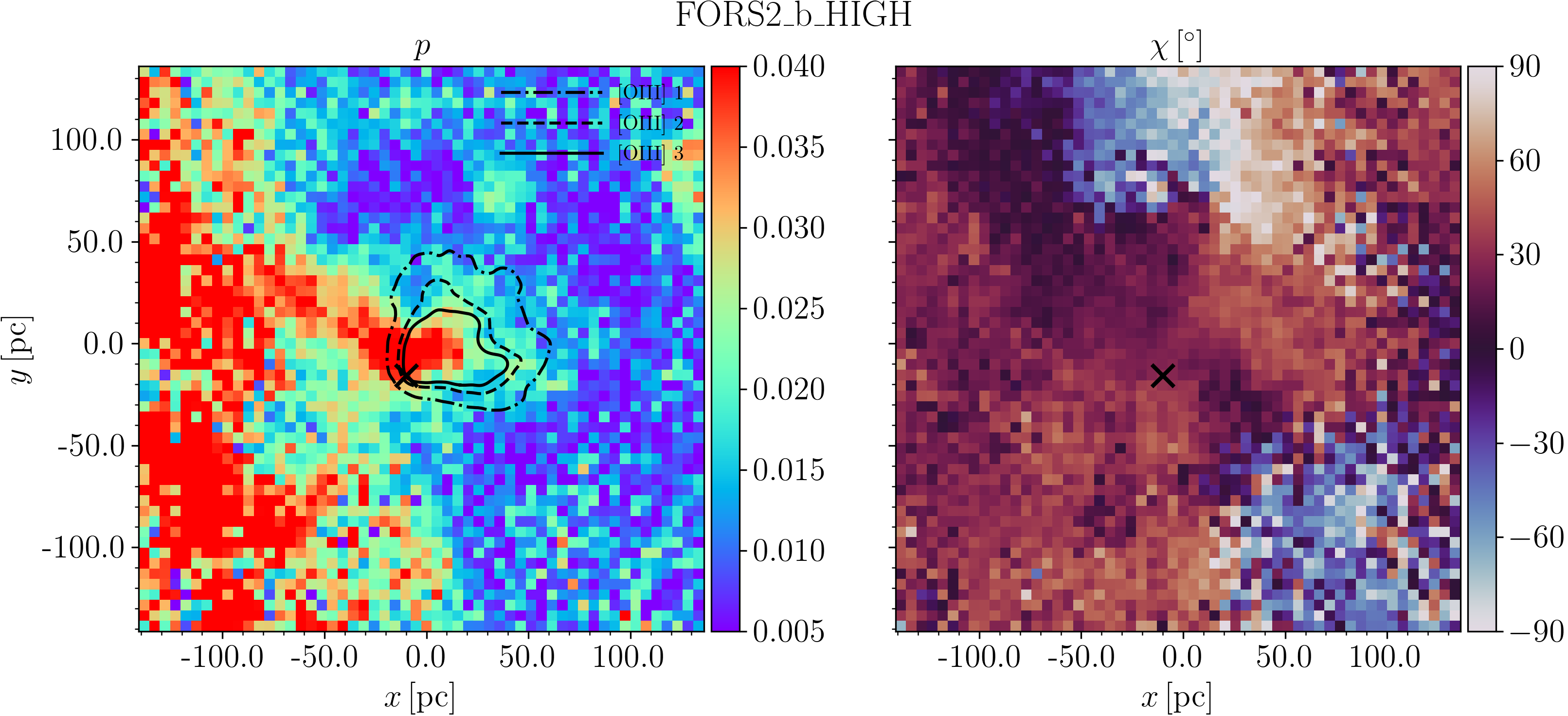}
 \includegraphics[width=0.49\textwidth]{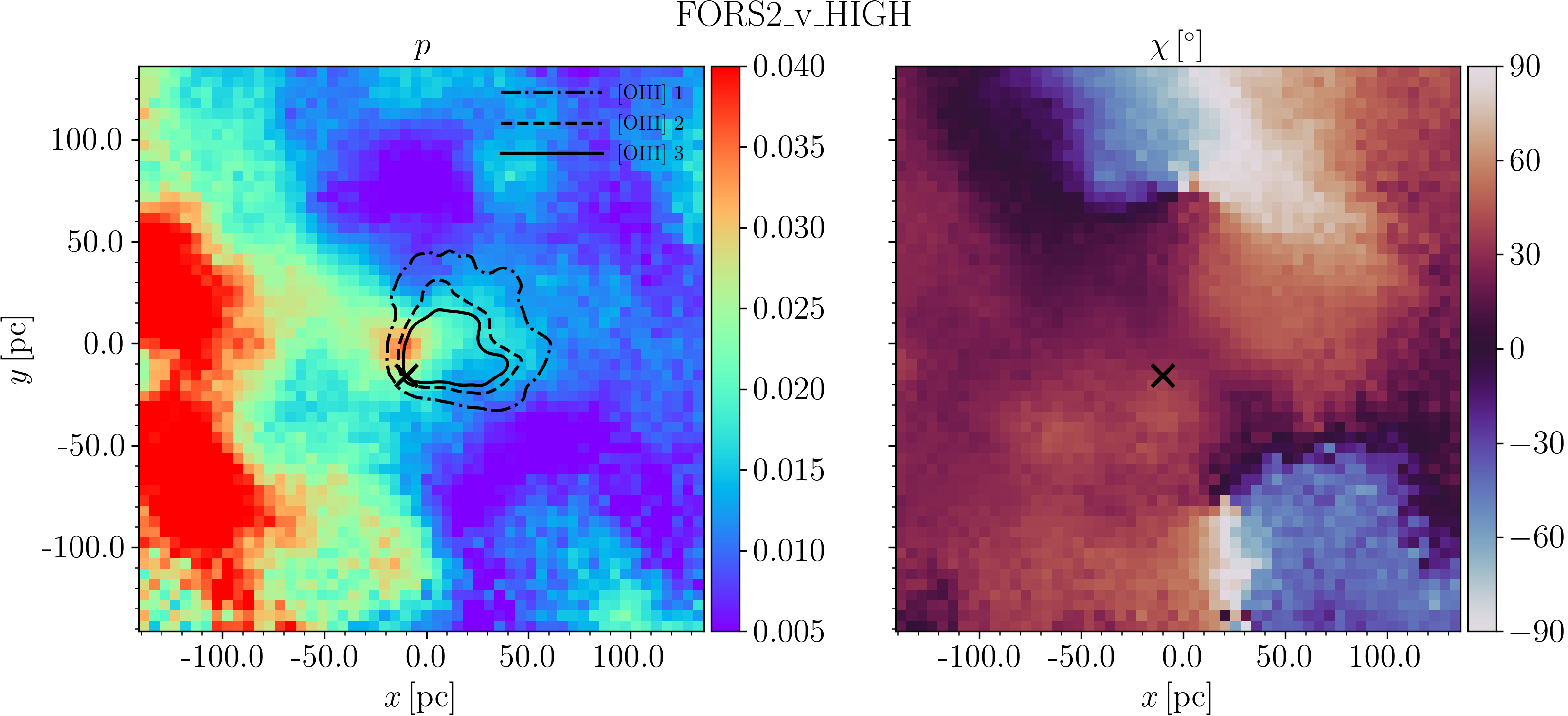}\\
 \includegraphics[width=0.49\textwidth]{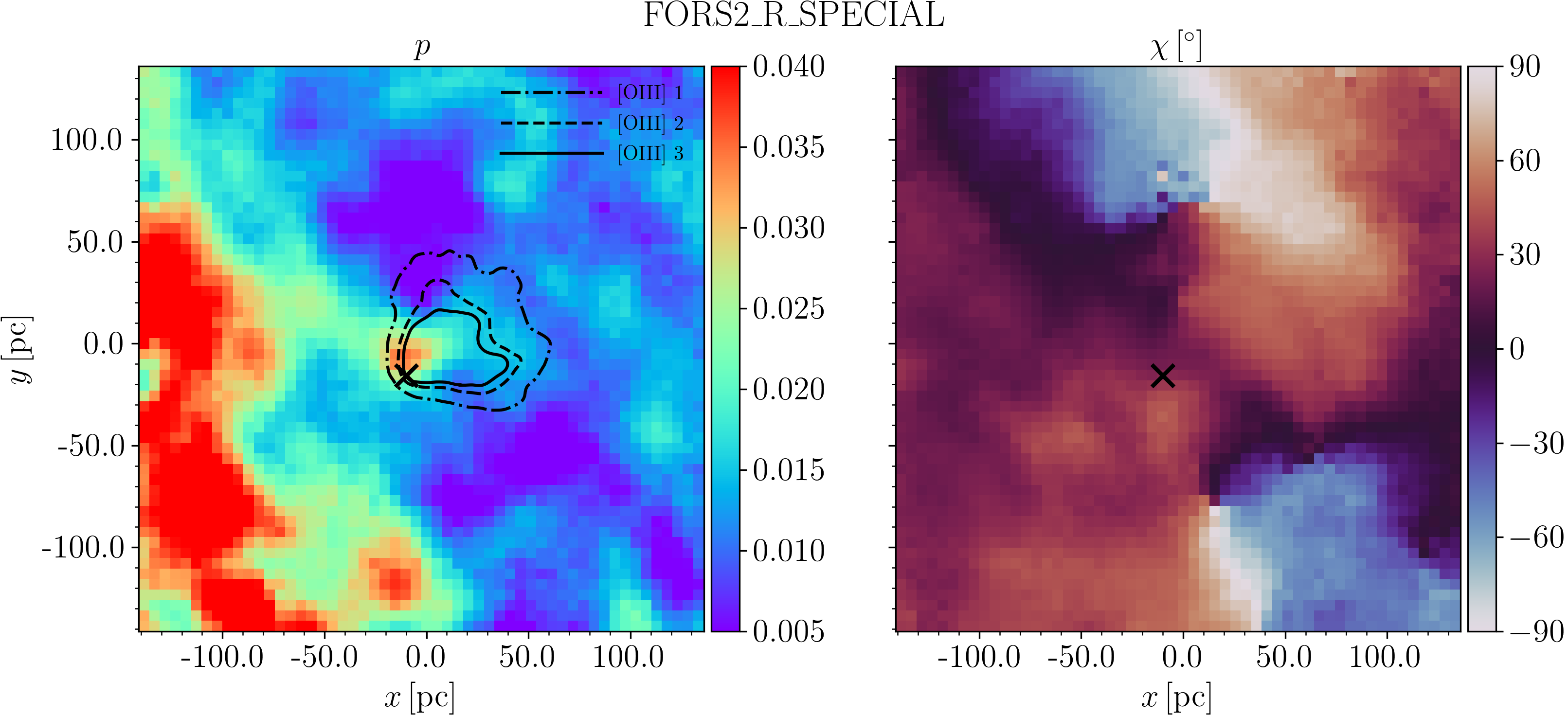}
 \includegraphics[width=0.49\textwidth]{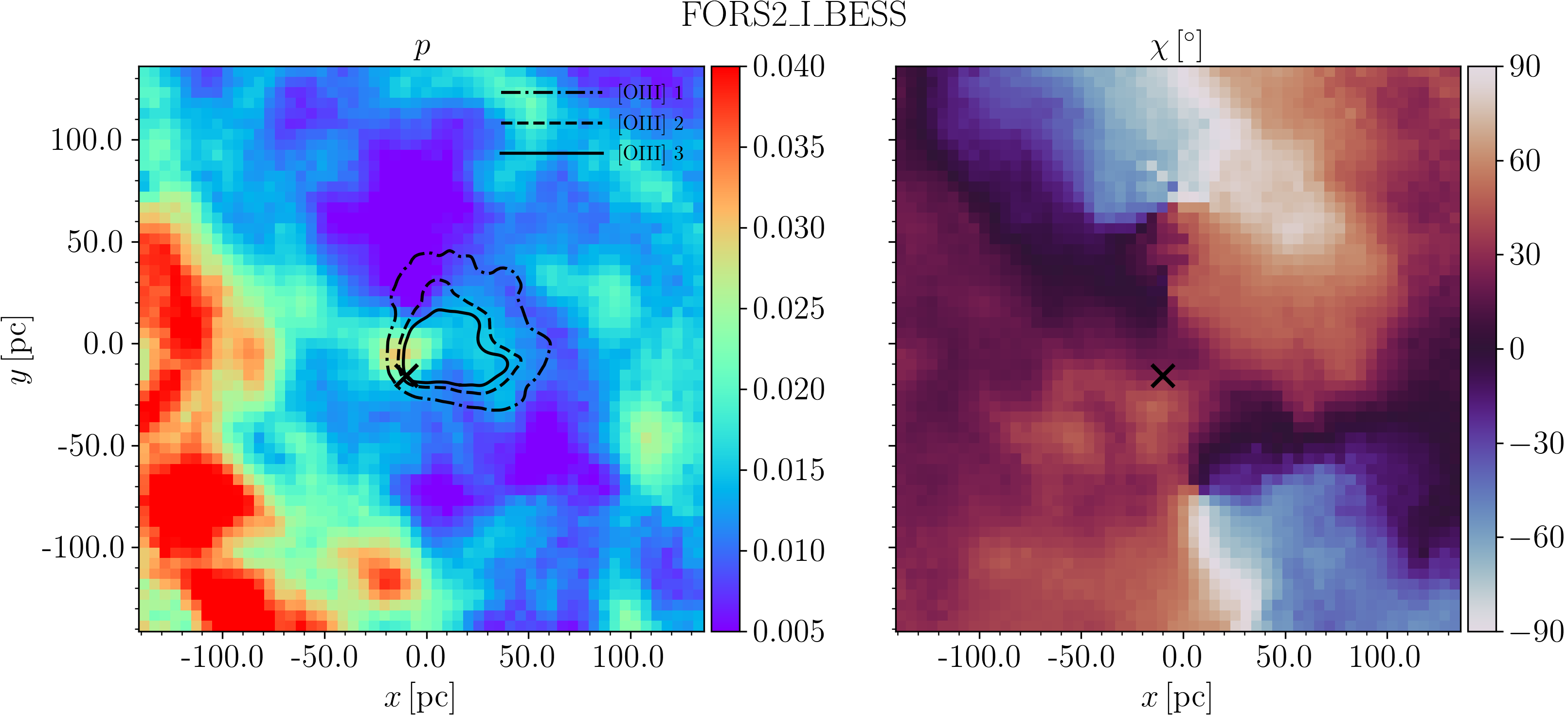}
 \caption{Polarization degree and angle maps of the central $280\times280$ ${\rm pc}^2$ obtained in the $BVRI$ filters. $B$-band maps are of notable lower quality due to being obtained from a single offset. Polarization degree maps clearly feature a conical structure that overlaps with the ionization cone, as indicated by the [\ion{O}{iii}] contours. X marks the position of the AGN nucleus.}
 \label{fig:obsPolMap}
\end{figure*}

\begin{figure*}
 \includegraphics[width=0.49\textwidth]{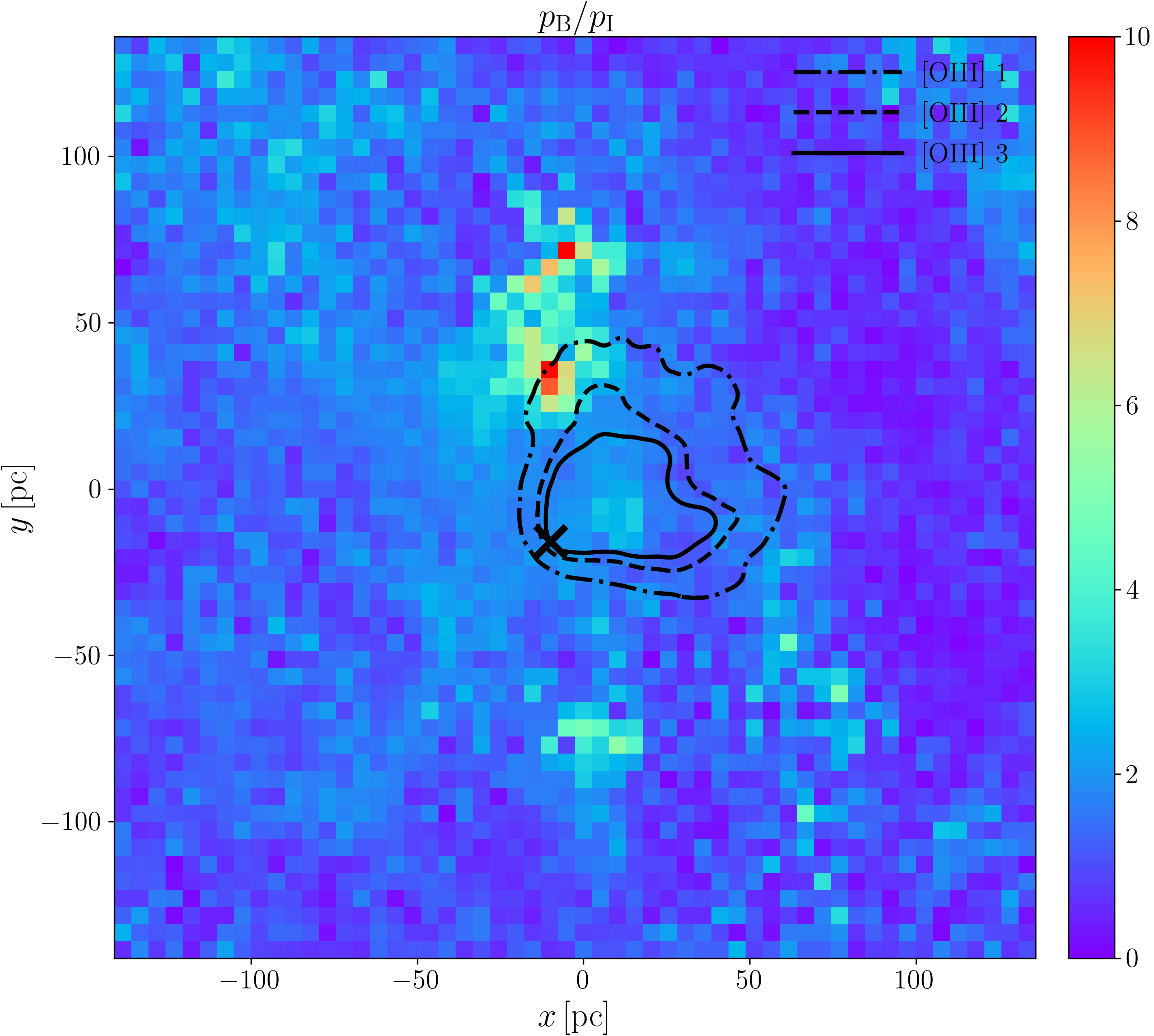}
 \includegraphics[width=0.49\textwidth]{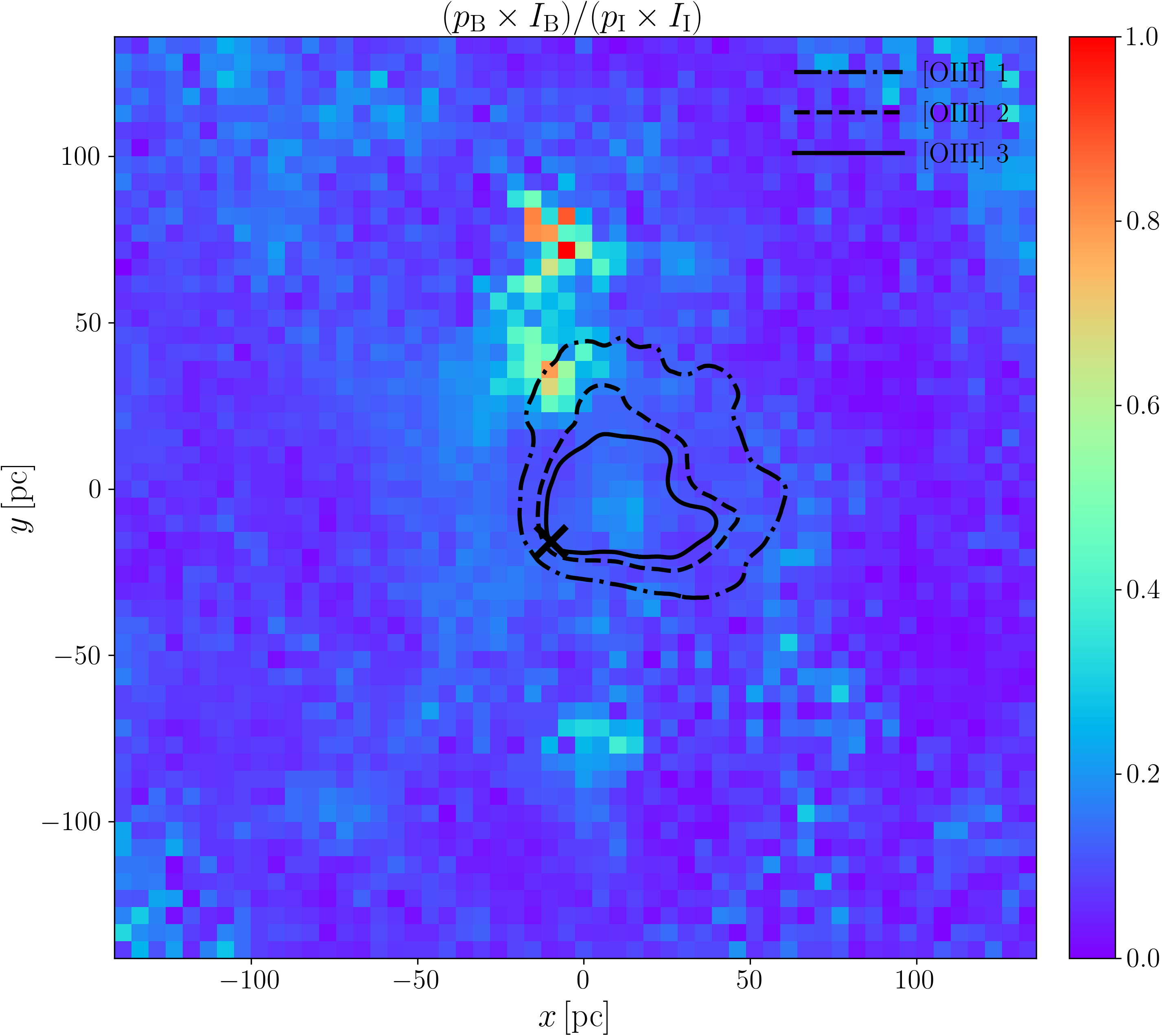}
 \caption{$B$-to-$I$ band ratio of the polarization degrees (left) and polarized fluxes (right) with the [\ion{O}{iii}] contours overploted. Higher values are indicative of larger contribution of the scattering on dust grains.}
 \label{fig:obsRatio}
\end{figure*}

The analysis of the entire host galaxy polarization and dust properties will be presented in a follow-up publication. 
In this work, our focus is on the AGN core and ionization cone, up to 40 pc ($\approx2\arcsec$) from the central engine. More precisely, we compare the prediction of the radiative transfer models to the data within the smallest and middle [\ion{O}{iii}] contours from VLT Multi-Unit Spectroscopic Explorer (MUSE) observations (Kakkad et al, subm.) overplotted in the Figs.~\ref{fig:obsPolVec},\ref{fig:obsPolMap},\ref{fig:obsRatio} with solid and dashed lines.  However, in the figures we present a region of  $280\times280$ ${\rm pc}^2$ to provide a wider context and estimate if and how the galaxy structures affect the observables. 
In Fig.~\ref{fig:obsPolVec} we show the intensity maps in the $BVRI$ filters\footnote{In the official FORS2 nomenclature: b\_HIGH, v\_HIGH, R\_SPECIAL, I\_BESS}. In the upper panels, vectors represent polarization degree, while in the lower panels they stand for the polarized flux given by: $p$ $\times$ $I$, where $I$ is the total intensity in the given band. The white circle in the top left corner of each panel represents median the FWHM of the PSF measured from the brightest stars in the field (see table \ref{tab:obs} for exact values).  Polarization induced by scattering of the photons originating in the central point source should result in a centro-symmetric pattern around the nucleus \citep{Kishimoto1999}. We see that the polarization vectors in the outer regions of our maps deviate from such a pattern, likely due to the spiral-like polarization features of the host galaxy: namely, if the dust grains are aligned, dichroic extinction may determine the orientation of the polarization angle, depending on the optical depth and magnitude of the alignment \citep{Kartje1995}. Fig.~\ref{fig:obsQU} features $Q$ and $U$ Stokes parameters maps of the same region and Fig.~\ref{fig:obsPolMap} shows the corresponding polarization degree and angle maps. The region of the highest polarization degree and lowest flux is associated with the structures of the host galaxy disc, which extends towards the observer in the lower left sections of the panels. On the other hand, the highest polarized flux comes from the region around the AGN nucleus and ionization cone. In the same area, coinciding with the [\ion{O}{iii}] MUSE contours observations, the polarization degree maps reveal a conical structure.
The integrated values within the cone (Table \ref{tab:con}) used to compare the observations with the models contain the average propagated error in every pixel (see Appendix \ref{app:stokes}), but also the standard deviation added in quadrature. This is a conservative approach and even so, the uncertainties of polarization are one order of magnitude lower (except for $B$-band that was not dithered).

\begin{table*}
\centering
\caption{Polarization degree and position angle within the ionization cone (obtained from the integrated Stokes $Q$ and $U$) defined by the three VLT/MUSE [\ion{O}{iii}] contours shown in Fig.~\ref{fig:obsPolVec}.}
\begin{tabular}{c|cc|cc|cc}
\hline
\hline
Filter & \multicolumn{2}{c}{contour \#3} & \multicolumn{2}{c}{contour \#2} & \multicolumn{2}{c}{contour \#1} \\
         & $p$(\%)       & $\chi$(deg)   & $p$(\%)       & $\chi$(deg)   & $p$(\%)       & $\chi$(deg)   \\
\hline
$B$      & 3.00$\pm$1.38 & 28.0$\pm$13.1 & 2.29$\pm$1.40 & 28.3$\pm$17.5 & 1.90$\pm$1.44 & 30.0$\pm$21.9 \\
$V$      & 1.94$\pm$0.33 & 32.9$\pm$4.9  & 1.75$\pm$0.33 & 33.6$\pm$5.4 & 1.49$\pm$0.35 & 33.5$\pm$6.8  \\
$R$      & 1.61$\pm$0.18 & 32.1$\pm$3.1  & 1.47$\pm$0.19 & 31.9$\pm$3.6 & 1.19$\pm$0.20 & 31.8$\pm$4.9  \\
$I$      & 1.49$\pm$0.22 & 31.3$\pm$4.3  & 1.35$\pm$0.23 & 32.0$\pm$4.8 & 1.05$\pm$0.24 & 33.6$\pm$6.5 
\end{tabular}
\label{tab:con}
\end{table*}

In the Fig.~\ref{fig:obsRatio}, we show the $B$-to-$I$ band ratio of the polarization degree and polarized flux. These maps do not show distinct morphology, but the region within and around the ionization cone does have higher values (compared to the darker stripes farther out), indicative of polarization by scattering on the dust grains, since scattering on electrons results in wavelength-independent polarization. A few "hot" pixels do not correspond  to any structure and could be just a consequence of low signal-to-noise of the $B$-band image.

Furthermore, the Stokes $U$ and polarization angle maps reveal a striking biconical morphology with a possible hint of a counter-cone hidden below the host galaxy disc. By definition, $+Q/-Q$ represent the component polarized at $PA=0 / 90\degr$ (North-South / East-West directions), while $+U/-U$ represent $45 / 135\degr$ direction. In other words, $Q$ describes whether the plane of polarization is observed to be vertical or horizontal, while $U$ describes whether the plane of polarization is diagonal in the image plane. The Circinus ionization cone is polarized at $PA\approx30-40\degr$, while the disc plane is polarized close to $PA\approx0\degr$. Thus, by coincidence, the ionization cone region and host disk region tend to reside in the $U$ and $Q$ parameters, respectively.

We conclude that dichroic absorption is likely responsible for the polarization pattern observed in the outer regions of presented maps, but there is sufficient evidence that, in the region of our interest (ionization cone, up to $\approx40$ pc from the nucleus) scattering on the dust grains is the dominant polarization mechanism, which we will compare to the models. Furthermore, the ionization cone is above the host galaxy disc; if we consider the region of the galaxy, which is in projection behind the cone, any light polarized by another mechanism (e.g. dichroism) will pass through the dusty cone on the way to the observer and will be partially absorbed or scattered away, reducing its contribution to the observed polarization. Finally, the presence of broad lines in the polarized spectrum of $1\arcsec\times1\arcsec$ aperture \citep{RamosAlmeida2016} is a reliable indicator of scattering being the dominant polarization mechanism in the considered region \citep[e.g][]{Antonucci1983}.

\subsubsection{Comparison to previous works}

\begin{figure*}
 \includegraphics[width=0.98\textwidth]{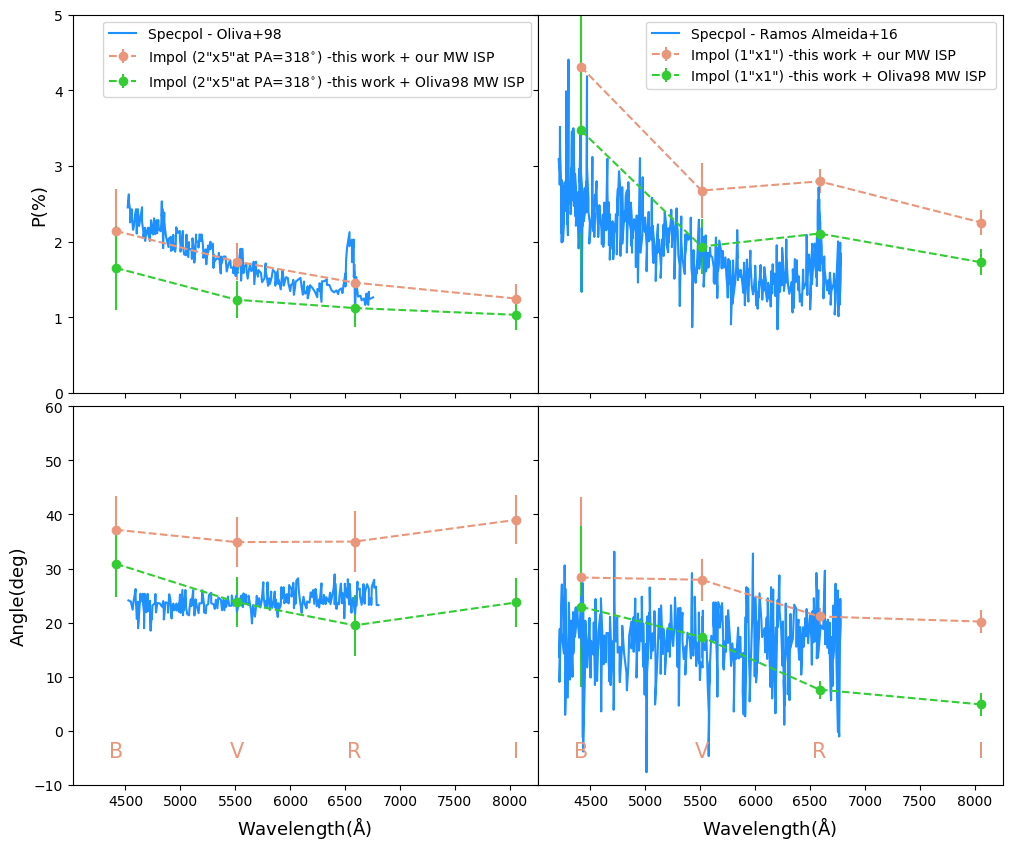}
 \caption{Imaging polarimetry from this work compared to literature spectro-polarimetry: \emph{Left:} MW-corrected EFOSC1/ESO3.6m optical spectro-polarimetry from \citet{Oliva98} with a $2\arcsec\times5\arcsec$ slit tilted by PA$=318\deg$ (blue) compared to the integrated imaging polarimetry in $BVRI$ bands from this work using our own MW ISP correction (orange) and with their MW correction (green). \emph{Right:} MW-corrected spectro-polarimetry taken with a $1\arcsec\times1\arcsec$ slit with FORS2/VLT from \citet{RamosAlmeida2016} (blue) compared to our imaging polarimetry from the central arcsec box aperture with our own MW correction (orange) and with their MW correction (green, taken from \citet{Oliva98}).}
 \label{fig:RA16}
\end{figure*}

In this section we compare our results with previous polarimetric observations of Circinus. There have been three previous studies of spectro-polarimetry of the Circinus nucleus: with EFOSC1 at the ESO-3.6m telescope \citep{Oliva98}, with a spectropolarimeter at the 3.9m Anglo-Australian Telescope \citep{Alexander2000} and with FORS2 at VLT \citep{RamosAlmeida2016}\footnote{Data extracted using: \url{https://automeris.io/WebPlotDigitizer/}.}. The overall polarization properties presented by \citet{Alexander2000} are very similar to \citet{Oliva98}, but are based on the lower signal-to-noise data. Thus, we will consider only the observations by \citet{Oliva98} and \citet{RamosAlmeida2016}. Both cover a similar spectral range but the slit had a different size and position. The former used a $2\arcsec$ slit that covered $5\arcsec$ in length tilted at 318$^{\circ}$, aligned with the direction perpendicular to the Circinus disk, whereas the latter used a $1\arcsec$ slit with a $1\arcsec$ aperture oriented with the parallactic angle. In Fig.~\ref{fig:RA16} we compare those observations with our imaging polarimetry integrated in the appropriate apertures abd re-oriented when necessary. Both the data of \citet{Oliva98} and of \citet{RamosAlmeida2016} were corrected for the MW ISP obtained from a single star in a single filter and extrapolated to other wavelengths using a standard Serkowski law. This correction was adopted from \citet{Oliva98}. Our MW ISP is more robust, as it is obtained from a large sample ($\approx800$) of field stars observed at multiple offsets and filters (see section~\ref{subsec:MW-ISP}). Nevertheless, for the sake of consistent comparison, we also show our values corrected with their MW ISP. 
It is evident that our measurements of polarization degree and angle are mostly higher that those inferred by \citet{RamosAlmeida2016} and \citet{Oliva98}.
However, when using the same MW ISP correction, the measurements agree within uncertainties with both literature spectro-polarimetric datasets of Circinus, with exception of the polarization degree, which provides a better match to \citet{Oliva98} with our correction. The difference between the inferred values with the two correction illustrates the importance of using multiple fields stars to determine ISP whenever possible.

\section{Radiative transfer models}
\label{sec:mod}

All the details of the models and an in-depth motivation for particular choices are discussed in \citetalias{Stalevski2017} and \citetalias{Stalevski2019}. We invite the reader to examine the relevant sections in those papers; here we lay out only the basic properties and point out some differences.

\subsection{Geometries}

The geometry of our fiducial model for the Circinus AGN is a composite of the two models we tailored to match the MIR observations on the scales of tens of parsecs \citepalias[VISIR,][]{Stalevski2017} and several parsecs \citepalias[MIDI,][]{Stalevski2019}. It consists of two components: a compact flared disc and a large scale hyperboloid shell resembling a hollow cone. The disc has a radius of 3 pc, it is geometrically thin (angular width between its equatorial plane and edge is $5\degr$) and thick to its own MIR radiation (at $9.7$ $\micron$) in the radial direction ($\tau_{9.7}=15$). The hyperboloid shell forms a vertical wall which bends to effectively become a hollow cone (extending to 40 pc) with an half-opening angle of $40\degr$, matching the value obtained from kinematic modelling of the narrow line region \citep{Fischer2013}. This component is at the limit of being optically thick (in the $V$-band) in the radial direction ($\tau_{V}$=1). We refer the reader to the section 4.1.5 of \citetalias{Stalevski2019} for more details and caveats which occur when "stitching" together MIDI- and VISIR-based models of different scales. The model is seen almost edge-on, at the inclination of $86\degr$. The accretion disc is tilted to the side of the cone by $40\degr$, to match the orientation of the inner part of the warped maser disc \citep{Greenhill2003}.
The model was shown to be consistent with the recent observations using the second-generation MIR interferometer VLTI/MATISSE \citep{Isbell2022}. Furthermore, \citet{Venanzi2020} presented 3D numerical simulations based on a semi-analytical model for dusty gas around a supermassive black hole exposed to the radiation of the AGN and hot dust itself. When inputting parameters of the Circinus, their simulations replicate the hyperboloid shape of the wind, very closely matching the above described geometry of our radiative transfer model.

\begin{figure*}
 \includegraphics[width=\textwidth]{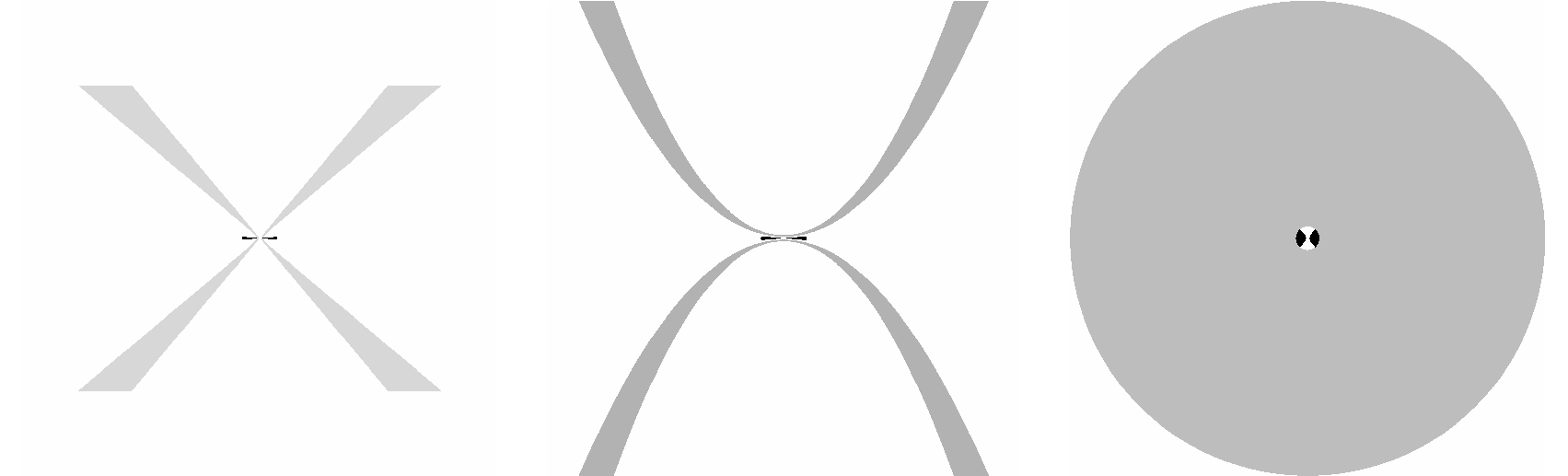}
 \caption{Sketch (density cut along the vertical plane) of the three model geometries we consider: disc$+$hyperboloid shell, disc$+$paraboloid shell and torus$+$spherical shell.}
 \label{fig:modGeo}
\end{figure*}

We also consider a geometry in which the hyperboloid cone component is replaced with a paraboloid shell with similar properties. Another geometry we model consists of a "torus" enveloped by a spherical shell. The torus is actually a flared disc, but with high covering factor, i.e. its angular width complements the opening angle of ionization cone. The spherical shell extends to 40 pc and has radial $\tau_{V}$=1. Our modelling in \citetalias{Stalevski2017} and \citetalias{Stalevski2019} strongly favoured hyperboloid shell geometry and excluded paraboloid and torus. However, we consider it is an useful exercise to examine polarimetric signatures of the different geometries to see if, in principle, they could be discerned by the observations. Fig.~\ref{fig:modGeo} shows a sketch of the three geometries. For brevity, hereafter, we will refer to them as the disc$+$hyp, disc$+$par, tor$+$sph.

\subsection{Dust properties}

For simplicity and due to the numerical limitations for the relatively large computational domain, dust in all the models is distributed homogeneously. The dust is composed of silicates ($53\%$) and graphite ($47\%$) in the disc components, and only graphite in the large scale components, with a typical Galactic grain size distribution $\propto a^{-3.5}$ \citep*[][]{MRN1977}, with sizes between $a_{\text{min}}=0.1$ and $a_{\text{max}}=1$ $\micron$. These choices are motivated by evidence for flat extinction curves in AGNs \citep{Gaskell2004, Shao2017, Xie2017} and theoretical expectation that the dust in the polar region should reflect the composition at its origin, i.e. the sublimation zone where only large graphite grains can survive \citep{Draine1984, Draine-Lee1984, Aitken1985, Barvainis1987, BaskinLaor2018}.

With high resolution spectroscopy of T-ReCS on Gemini-South, \citet{Roche2006} mapped the $9.7$ $\micron$ silicate feature depth across the polar region (the edges of the ionization cone). Inferred high values ($\tau_{9.7}\approx2$) indicate significant foreground absorption, likely due to the dust in the host galaxy disc. To account for this, when modelling MIR data of VISIR and MIDI, we applied an extinction screen on our models before comparing them to the observations. The same step should be taken in the present work. However, given its very high optical depth, the optical emission would be completely absorbed by such an uniform screen of dust. In reality, this foreground screen must be patchy or clumpy, allowing the partial transmission of the photons scattered in the dusty polar region. Here, we do not apply the extinction to the models, but the implications shall be discussed in Sec.~\ref{sec:modVsobs}.

\subsection{Radiative transfer simulations}

We employed \textsc{skirt}\footnote{\url{http://www.skirt.ugent.be}}, a state-of-the-art code based on the Monte Carlo radiative transfer (MCRT) \citep{Baes2011, Baes-Camps2015, Camps-Baes2015, Camps2020}. The code includes all the relevant physical mechanisms for processing the ultraviolet (UV) and optical radiation by scattering and absorption on dust grains and subsequent re-emission in the IR, and supports polarization by scattering on dust and electrons \citep{Peest2017}. To ensure optimal sampling and memory consumption, computational domain (a box with a side of 80 pc) is divided into a large number of small cells in a hierarchical adaptive octree grid \citep{Saftly2013} based on the dusty density of the underlying model geometry. We ran the simulations in eight wavelength bands covering most of the UV and optical range with $10^9$ photon packets emitted from the primary source. The central engine of an AGN (an accretion disc) was approximated by a point source with anisotropic emission that accounts for the change in the projected surface area and the limb-darkening as a function of the polar angle \citep[$L \propto \cos\theta(2\cos\theta+1)$,][]{Netzer1987}. Its SED was defined by a standard composition of power laws \citep{Stalevski2016}. For the bolometric luminosity of the primary source we adopted the value of $L_{\text{AGN}}=4\times10^{10} L_{\sun}$ \citepalias{Stalevski2017}. This is within the range inferred by different observations ($6\times10^{9}-7\times10^{10} L_{\odot}$, \citealt{Oliva1999, Arevalo2014, Tristram2014, Ricci2015}), but see \citetalias{Stalevski2019} for the discussion of possible discrepancy between the luminosity required by the small and large scale models.

\section{Results and discussion}
\label{sec:res}

\subsection{Polarization maps and signatures of the MCRT models}
\label{sec:modVsmod}

Pioneering work on polarization in AGN environment in the UV-optical range due to scattering on dust grains and electrons with MCRT approach has been done by \citet{Kartje1995}, \citet{Kishimoto1996} and \citet{Wolf-Henning1999}. More recently, the authors of the \textsc{stokes}\footnote{\url{http://astro.u-strasbg.fr/~marin/STOKES_web/index.html}} code  \citep{Goosmann2007}, have explored type 1 / type 2 dichotomy with different equatorial and polar scattering regions of smooth and fragmented media, including dilution by the interstellar polarization and host galaxy in a series of papers \citep{Marin2012,Marin2013,Marin2015,Marin2018}. During the implementation of the polarization in \textsc{skirt} \citep{Peest2017}, we have thoroughly tested the results with the \textsc{stokes} outputs and found good agreement between the two codes. Thus, we shall consider only the models tailored specifically for Circinus AGN based on our previous IR modelling \citepalias{Stalevski2017,Stalevski2019}.

\begin{figure*}
 \includegraphics[width=\textwidth]{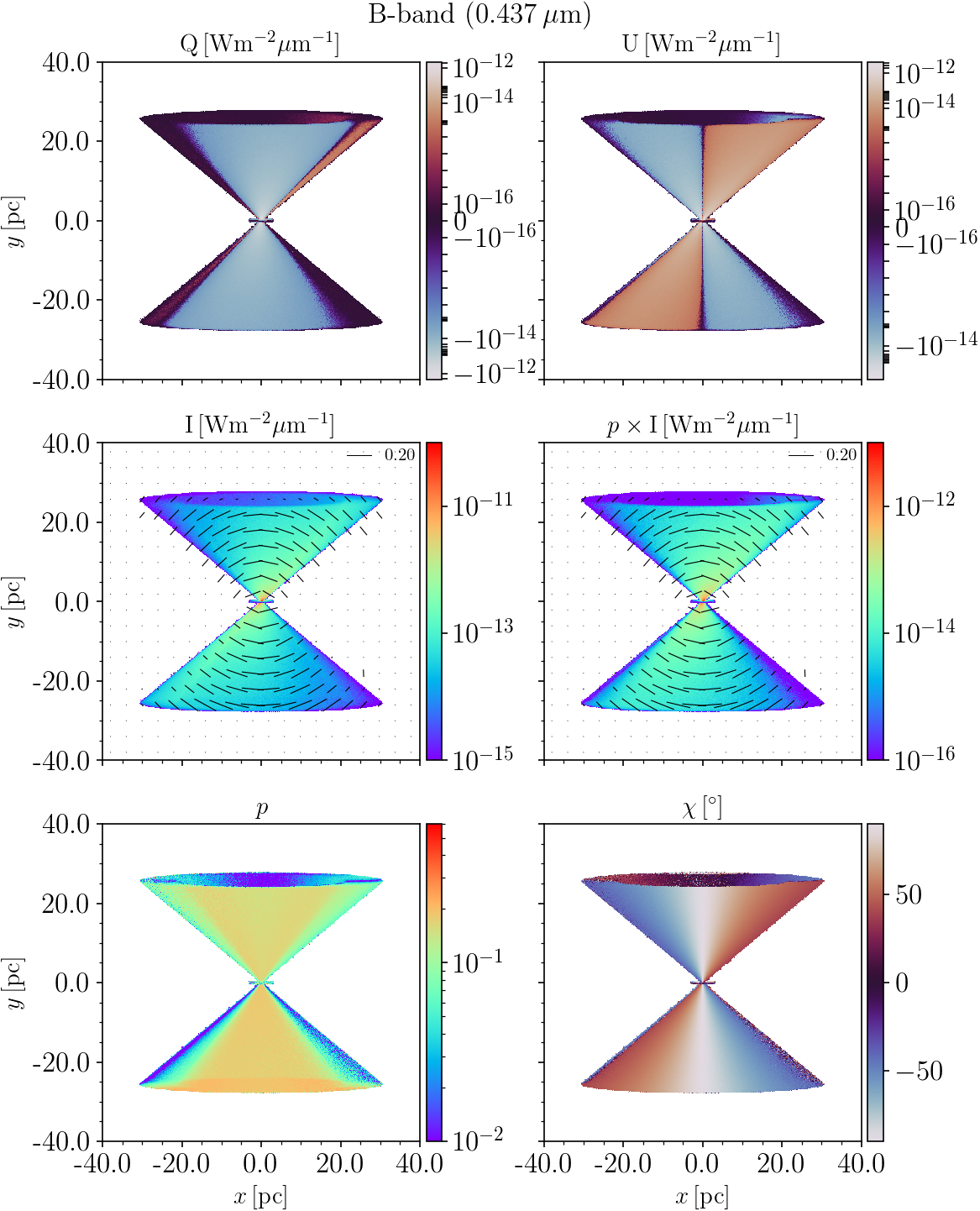}
 \caption{MCRT simulated $B$-band maps of the \texttt{disc+hyp} Circinus model at the viewing angle of $86\degr$: Stokes $Q$ and $U$ parameters (top row), polarization vectors overplotted on the intensity and polarized flux (middle row), polarization degree and angle (bottom row). The length of the vectors corresponds to the polarization degree (the scale indicated in the top right corner), their orientation corresponds to the polarization angle measured from the polar cone axis.}
 \label{fig:cirModFull}
\end{figure*}

\begin{figure*}
 \includegraphics[width=\textwidth]{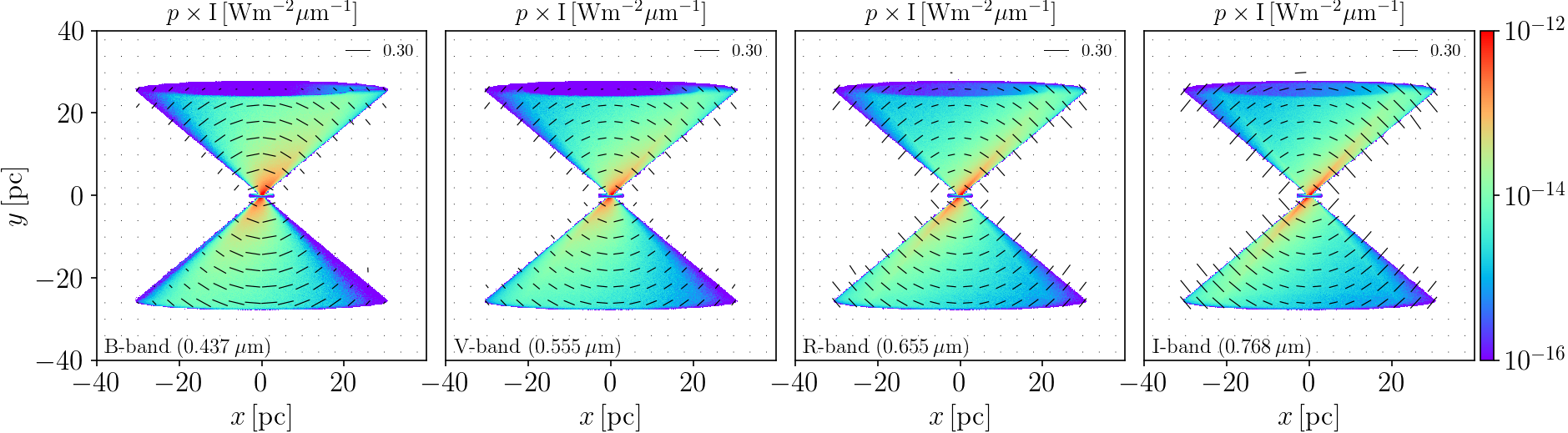}
 \caption{MCRT simulated $BVRI$ maps of the polarized flux with polarization vectors for the \texttt{disc+hyp} Circinus model at the viewing angle of $86\degr$. The length of the vectors corresponds to the polarization degree (the scale indicated in the top right corner), their orientation corresponds to the polarization angle measured from the polar cone axis.}
 \label{fig:cirModLambda}
\end{figure*}

In Fig.~\ref{fig:cirModFull} we showcase the $B$-band maps of the fiducial \texttt{disc+hyp} Circinus model: the Stokes parameters (top row), polarization vectors overplotted on the flux (middle left) and polarized flux (middle right) and polarization degree and angle (bottom row). Fig.~\ref{fig:cirModLambda} displays polarized flux maps with the polarization vectors in all four $BVRI$ bands. The polarization angle is represented with respect to the polar (cone) axis. To compare with the observations, the models should be rotated by $40\degr$ to match the on-sky orientation; this is not done in the displayed maps, but it is in all the subsequent plots which feature integrated values as a function of wavelength. The polarization degree of the \texttt{disc+hyp} model (Fig.~\ref{fig:cirModFull}, bottom left panel) is uniform in the central part of the cone and decreases towards the edges. However, this gradient is not likely to be detected in the observed maps due to the achieved angular resolution (see indicated PSF size in the Fig.~\ref{fig:obsPolVec}).

\begin{figure*}
 \includegraphics[width=0.825\textwidth]{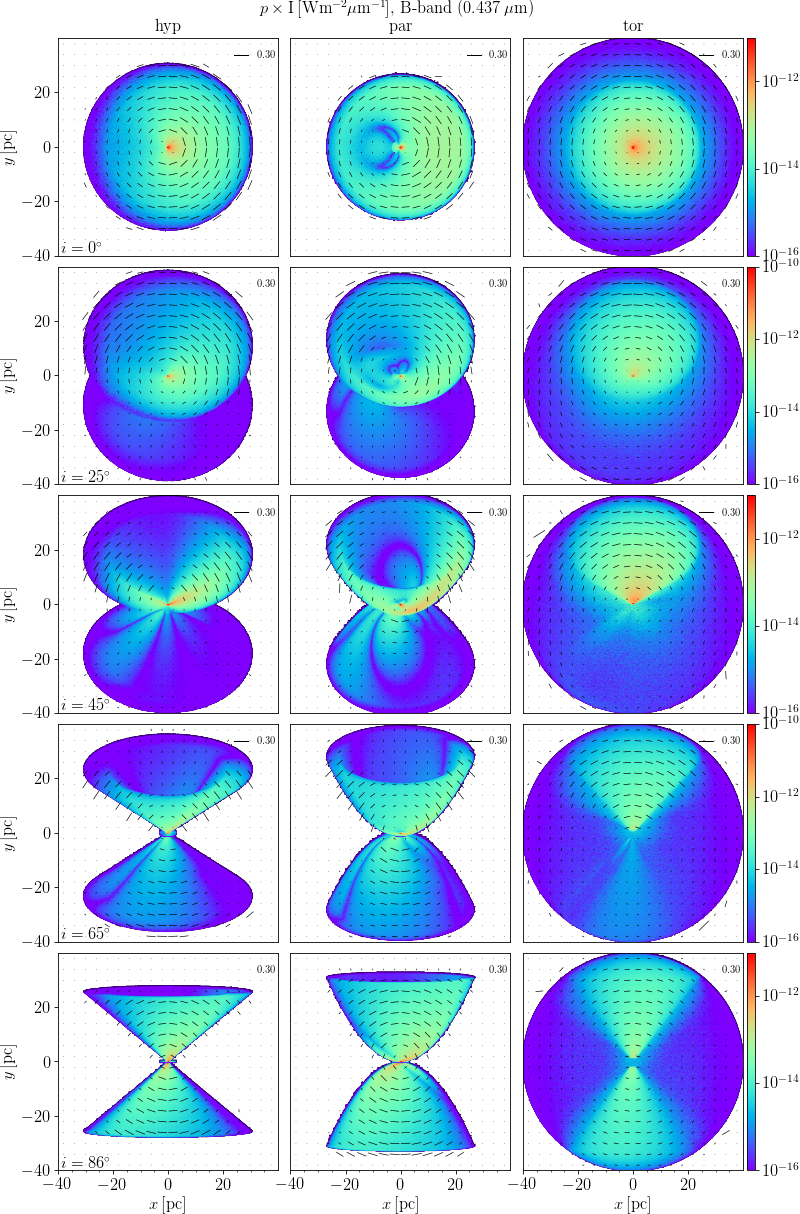}
 \caption{MCRT simulated $B$-band maps of the polarized flux with the polarization vectors for \texttt{disc+hyp}, \texttt{disc+par}, \texttt{tor+sph} models at different viewing angles, from face-on (top row) to edge-on (bottom row). The length of the vectors corresponds to the polarization degree (the scale indicated in the top right corner), their orientation to the polarization angle measured from the polar cone axis.}
 \label{fig:HypParTor}
\end{figure*}

Although our previous work of modeling the IR imaging, interferometric data and SED, together with constraints from the observed optical and maser emission, make a strong case for the \texttt{disc+hyp} model, we consider it a useful exercise to perform a limited parameter study to examine the polarimetric signatures of the different geometries and viewing angles that could characterize other objects. 
There are very few sources in which one could achieve comparable angular resolution and probe the spatial scales with polarimetry as it is the case in Circinus. In most cases, one could only study polarization integrated along a particular direction of the ionization cone, or even in the entire nucleus. Thus, results of MCRT simulations, such as presented here, are of crucial importance for interpretation of the observations.
In Fig.~\ref{fig:HypParTor} we display a grid of polarization maps of the three model geometries along a number of viewing angles, from face-on to edge-on. The asymmetries, seen most clearly in the polarized flux maps, are a consequence of the tilted anisotropic accretion disc. Some peculiar features, such as loops and streaks seen in the intermediate inclinations, are result of the illumination pattern by the anisotropic primary source and a particular geometry. Again, all these features are below the achieved angular resolution of Circinus FORS2 images -- they would be washed out when convolved with the PSF.

\begin{figure*}
 \includegraphics[width=0.49\textwidth]{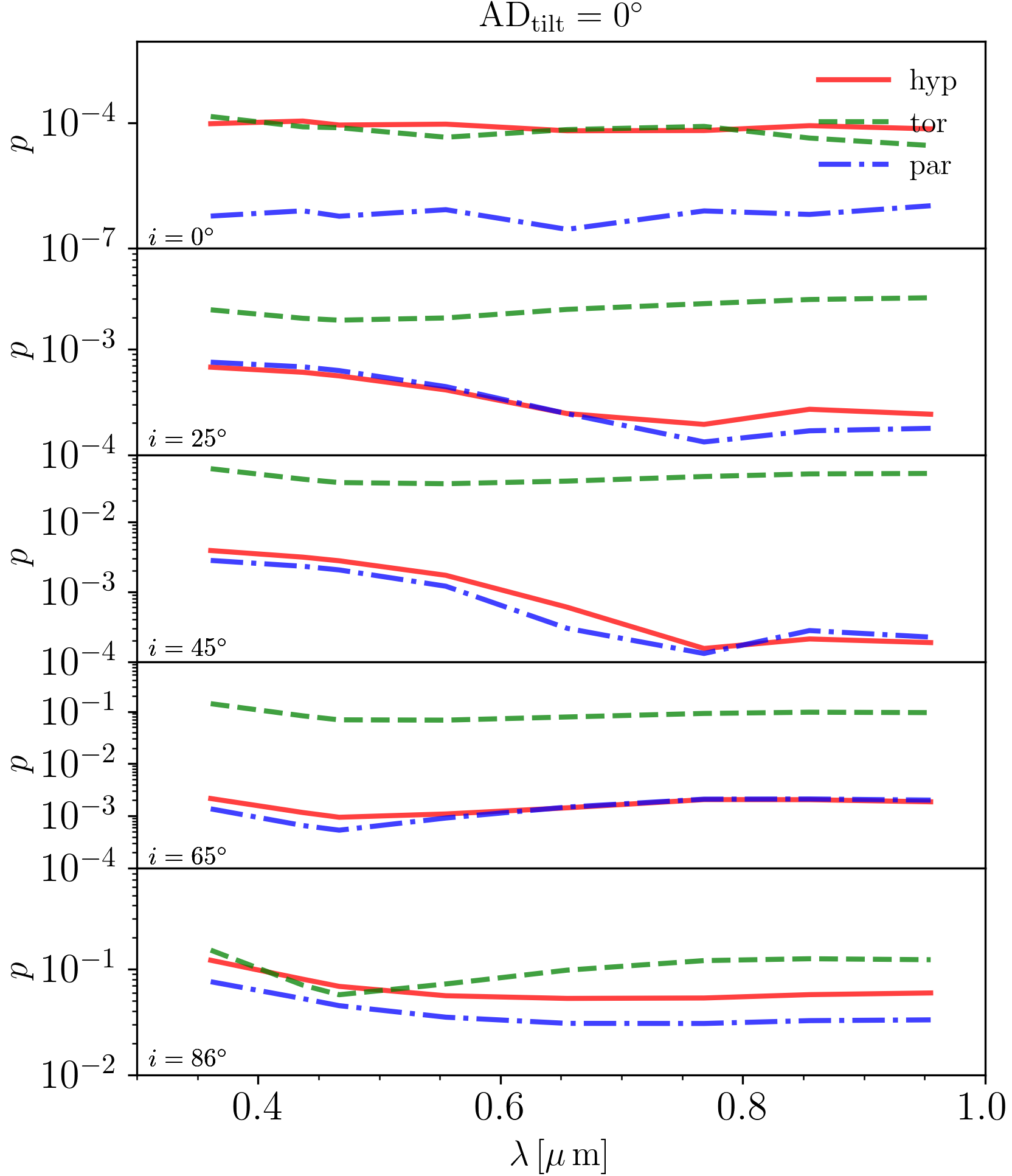}
 \includegraphics[width=0.49\textwidth]{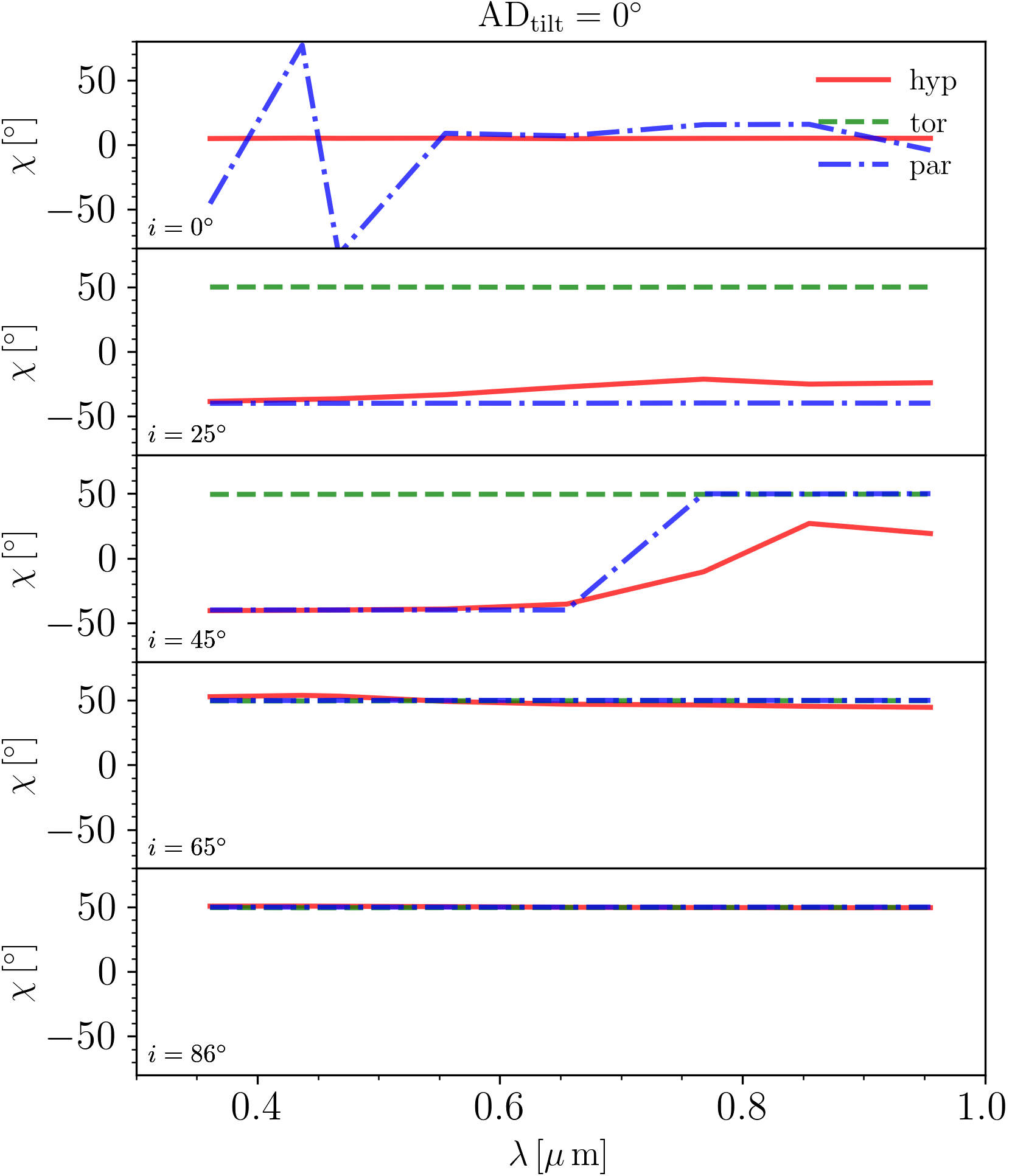}\\
 \includegraphics[width=0.49\textwidth]{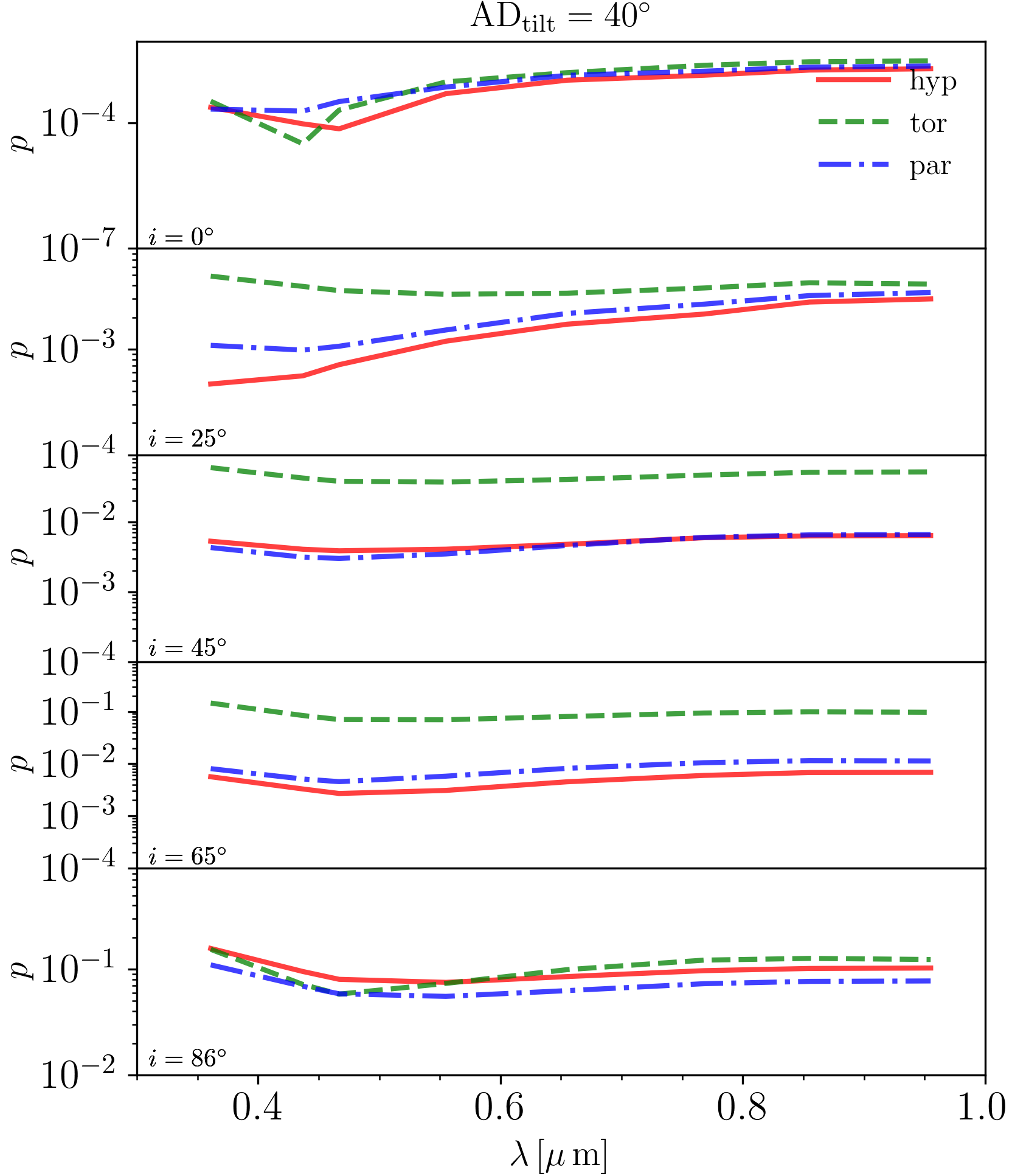}
 \includegraphics[width=0.49\textwidth]{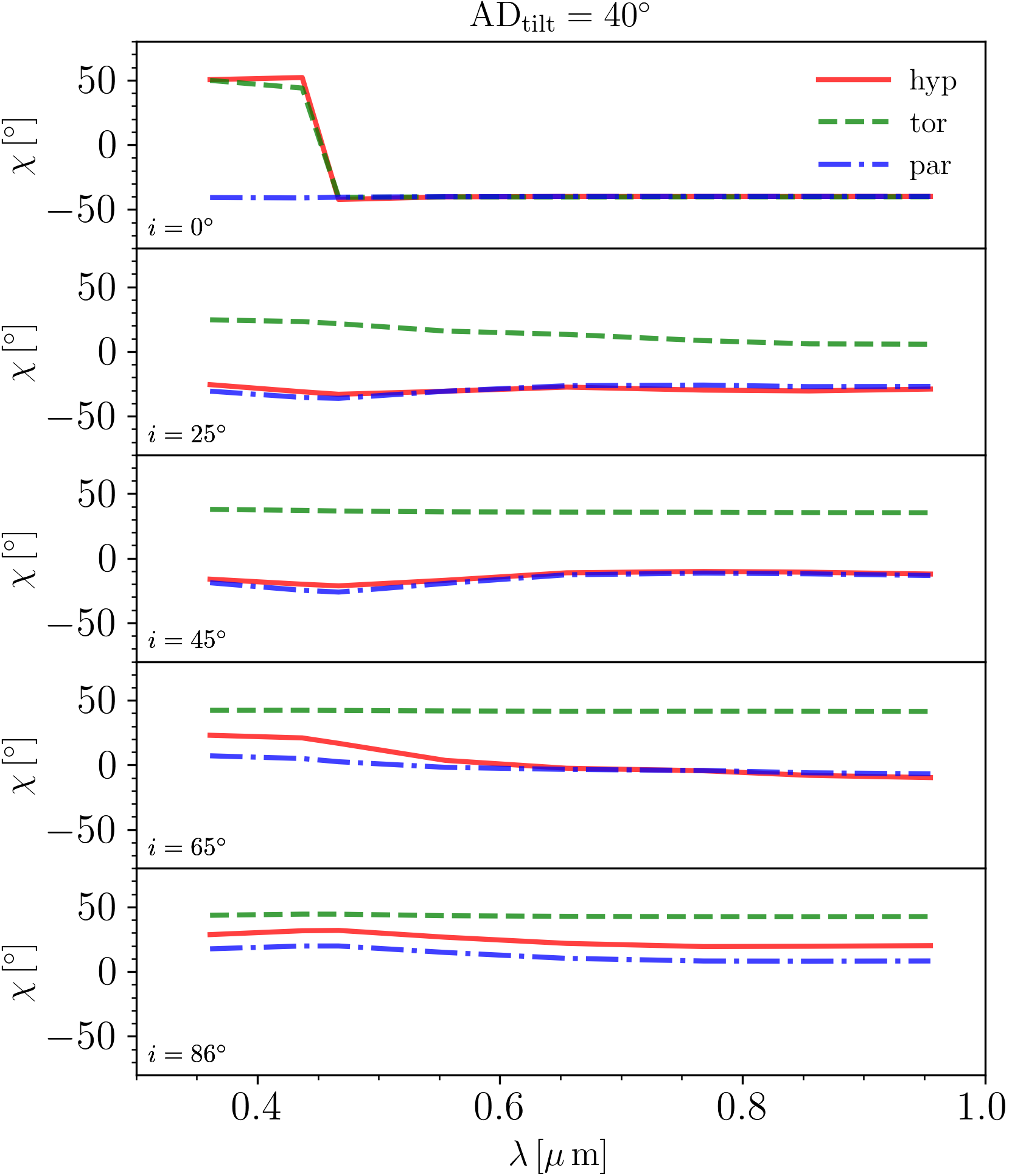}
 \caption{Polarization degree (left) and angle (right) integrated over the entire images of the models for five inclinations ($i=0, 25, 45, 65, 86\degr$) and for the accretion disc aligned with the cone axis (top) and tilted by $40\degr$ (bottom). Integration is preformed on Stokes $Q$ and $U$. Models \texttt{disc+hyp}, \texttt{disc+par}, \texttt{tor+sph} are labelled `hyp', `par', `tor' and shown in red solid, blue dash-dotted and green dashed lines, respectively. Prior to integration, the models have been rotated to match the on-sky orientation of Circinus AGN.}
 \label{fig:modIntTot}
\end{figure*}

\begin{figure*}
 \includegraphics[width=0.49\textwidth]{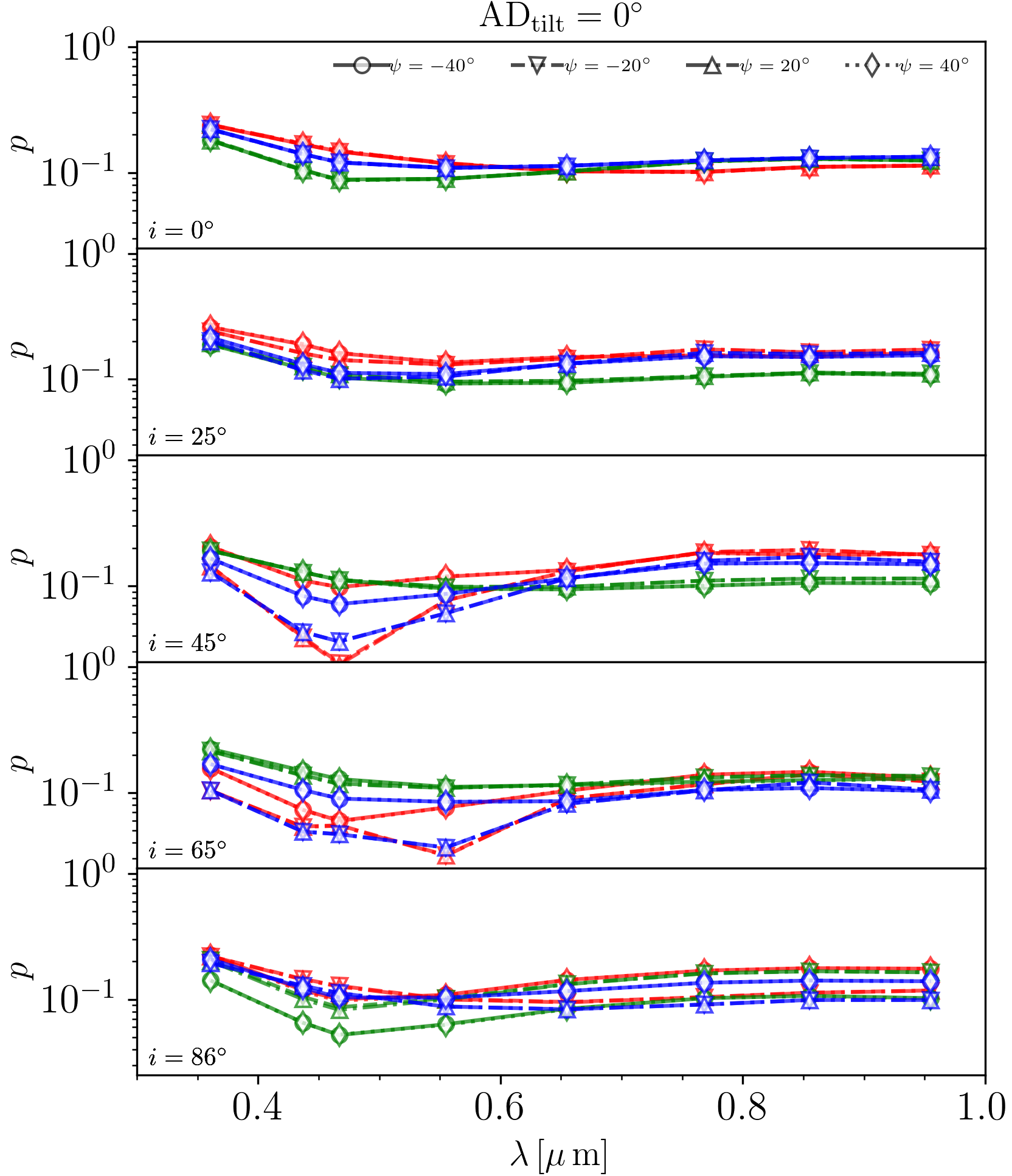}
 \includegraphics[width=0.49\textwidth]{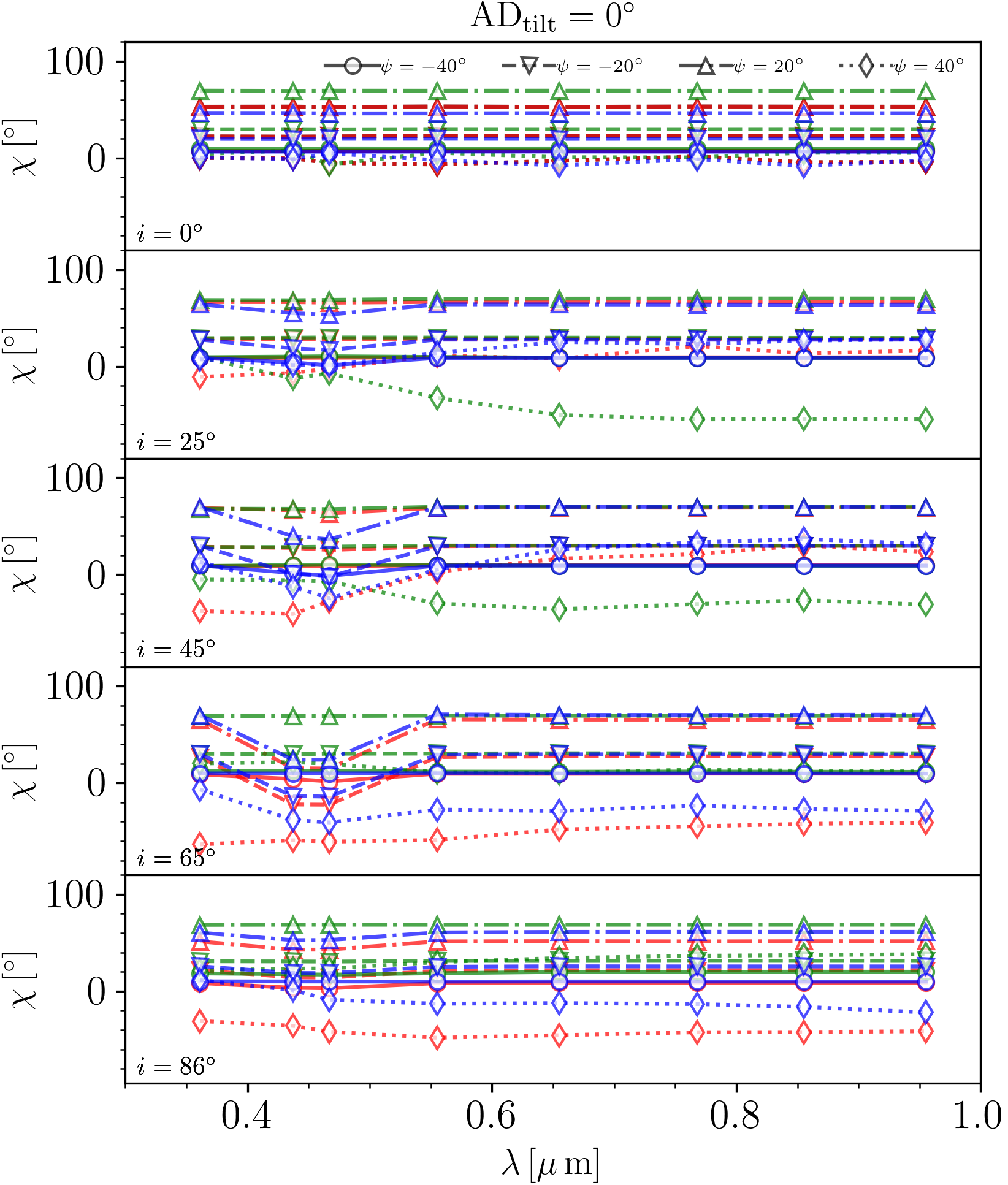}\\
 \includegraphics[width=0.49\textwidth]{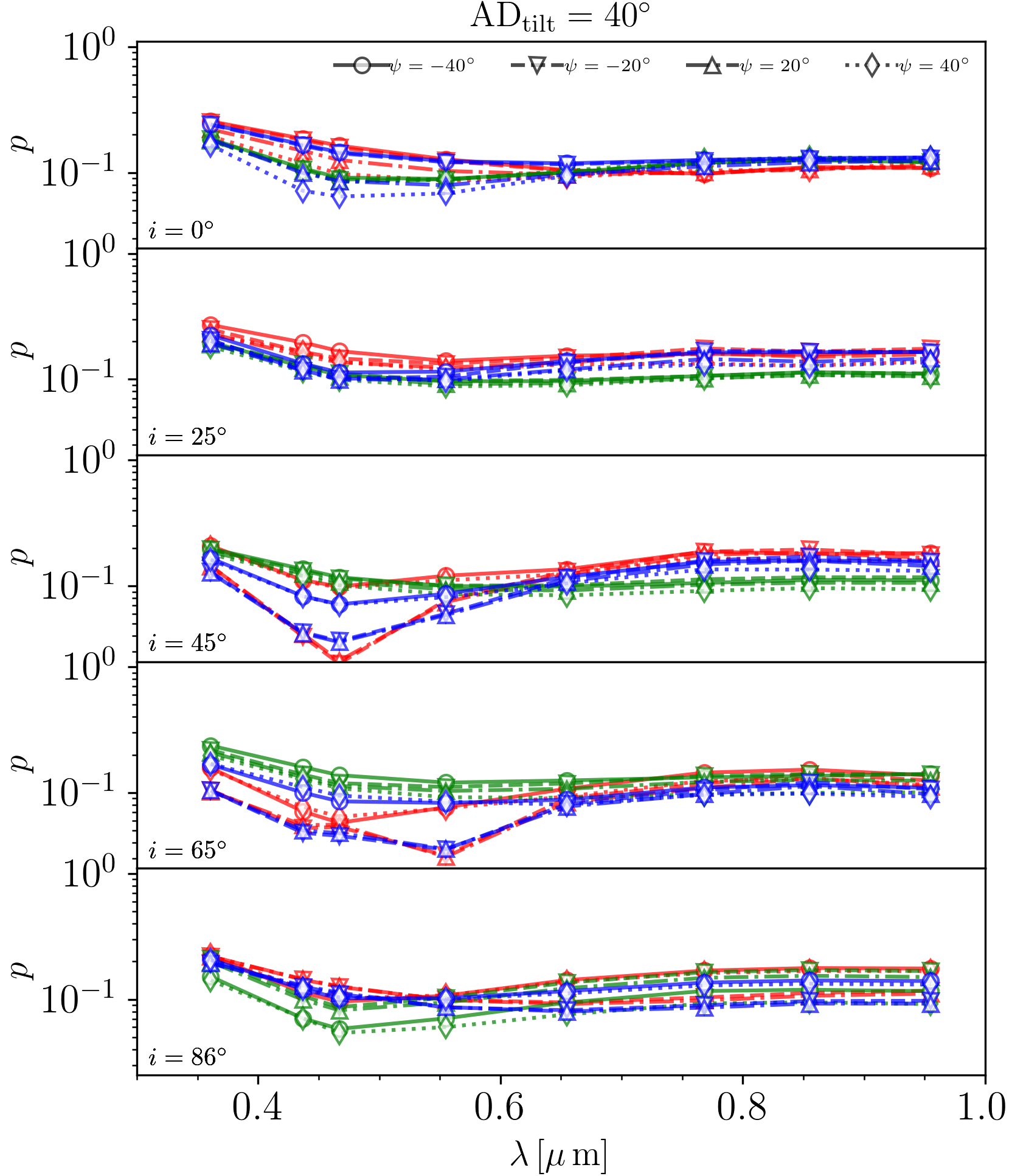}
 \includegraphics[width=0.49\textwidth]{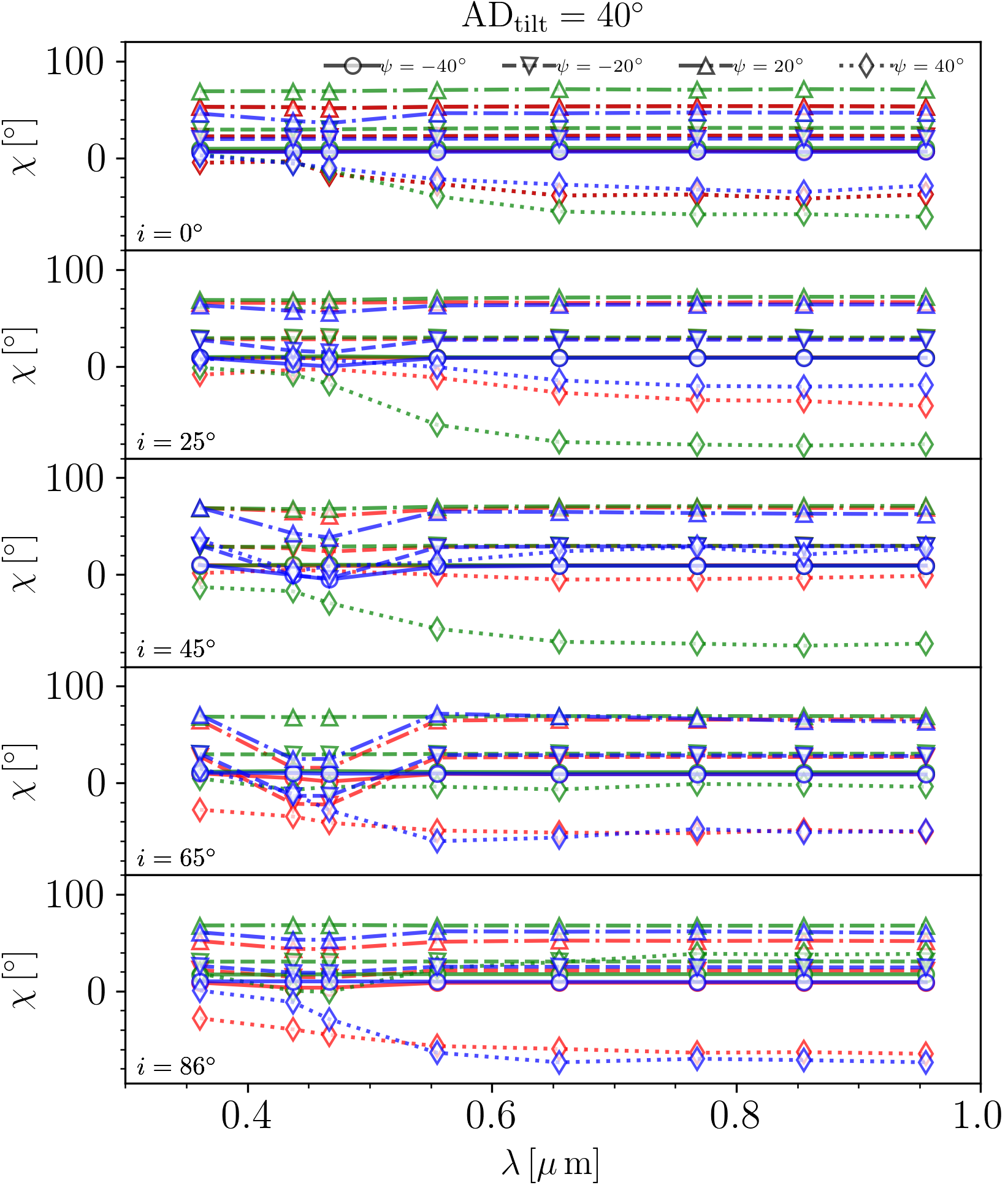}
 \caption{The same as in the Fig.\ref{fig:modIntTot}, but integrated along four directions (angle $\psi=\pm20, \pm40\degr$, with the reference direction being the cone axis). Models \texttt{disc+hyp}, \texttt{disc+par}, \texttt{tor+sph} are shown in red, blue and green lines. Different line types and symbols correspond to the directions of integration, is indicated in the top panels.}
 \label{fig:modIntDir}
\end{figure*}

In Fig.~\ref{fig:modIntTot} we present integrated values of polarization degree and angle as a function of wavelength for the entire model maps (integration is preformed on Stokes $Q$ and $U$) for the three model geometries and five viewing angles from face-one to edge-on. Similarly, in Fig.~\ref{fig:modIntDir} we present integrated values along the four directions across the maps. The purpose of this exercise is to examine if the various geometries could be distinguished when observing unresolved sources or for the different orientations of the spectropolarimetric slit. We consider two cases of the accretion disc orientations: aligned with the cone axis (\ADtilt{0}) and tilted towards the edge of the cone (\ADtilt{40} ). From these figures, we see that at longer wavelengths models show similar behaviour of polarization degree. Below $\approx0.6$ $\micron$ the models exhibit different wavelength dependence and, in principle, could be distinguished in some cases. Models \texttt{disc+hyp} and \texttt{disc+par} are in most cases similar and distinct from the \texttt{tor+sph}. Examining the polarization angle, in general, models are more clearly separated, both in fully integrated maps and for the various "slit" directions. Concerning the accretion disc orientation, it is interesting to note that for close to edge-on inclinations, the two tilt angles provide different predictions: for \ADtilt{0} all three models result in the same polarization angle, while for the \ADtilt{40} they are clearly separated (compare panels $i=65, 86\degr$ in the top and bottom panel of the right-hand side in Fig.~\ref{fig:modIntTot}). This demonstrates that polarimetry has a potential to probe not only the ISM, but also the properties of the primary source of illumination, such as its anisotropy and orientation. We shall examine this more closely in the next section.

\subsection{Polarization of Circinus AGN: models versus observations}
\label{sec:modVsobs}

We remind the reader that the extent of the models is $40$ pc ($\approx2\arcsec$) from the nucleus, coinciding with the narrow line region (as seen e.g. in [\ion{O}{iii}] emission), and thus, we limit the comparison to roughly this distance within the ionization cone. As argued in the Sec.~\ref{sec:obs-maps}, dominant polarization mechanism within this projected area should be scattering on dust grains. Regions farther out likely have significant contribution by the dichroic absorption and trace the host galaxy structures, which will be examined in a follow-up publication.

\begin{figure*}
 \includegraphics[width=0.49\textwidth]{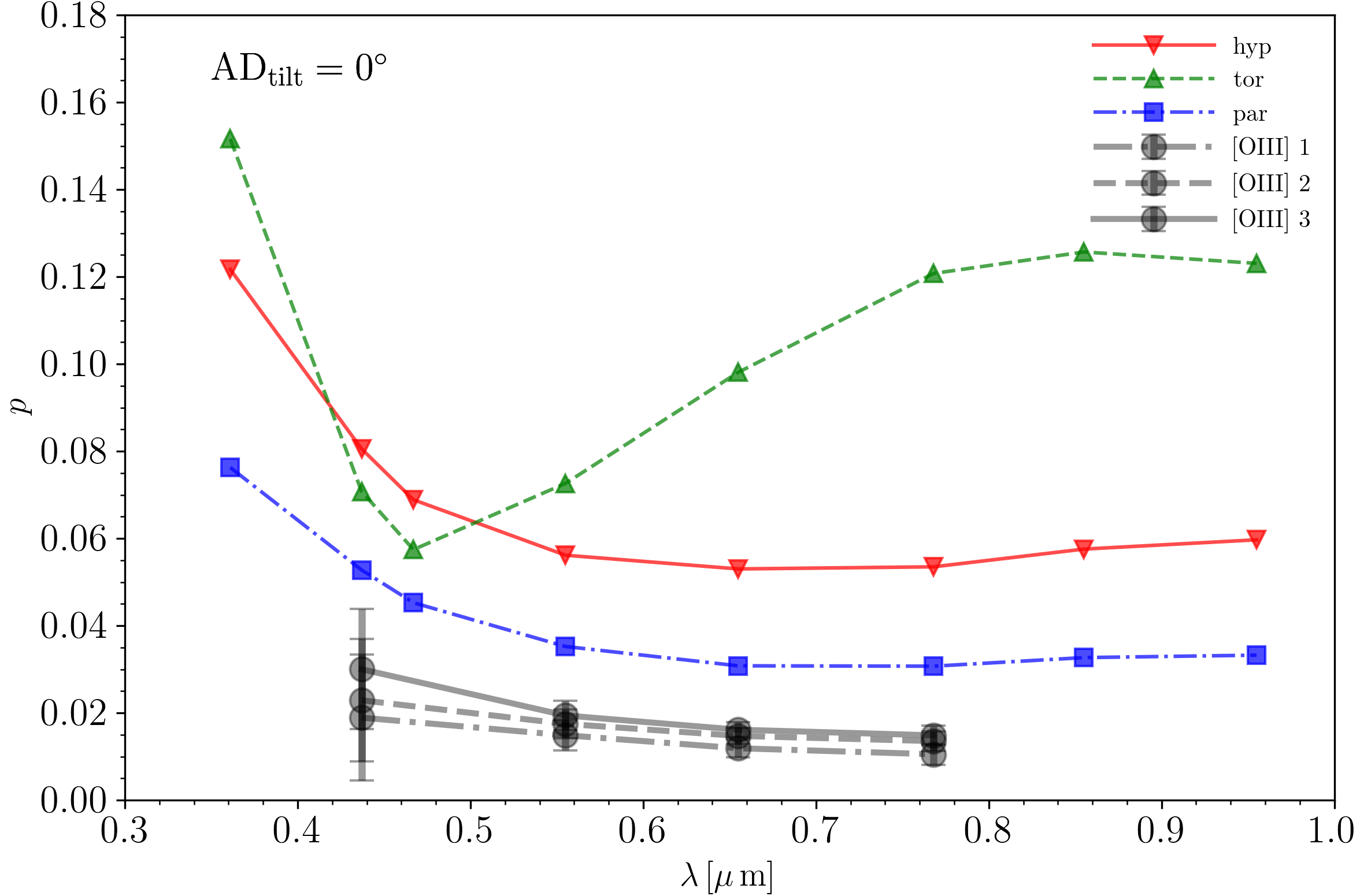}
 \includegraphics[width=0.49\textwidth]{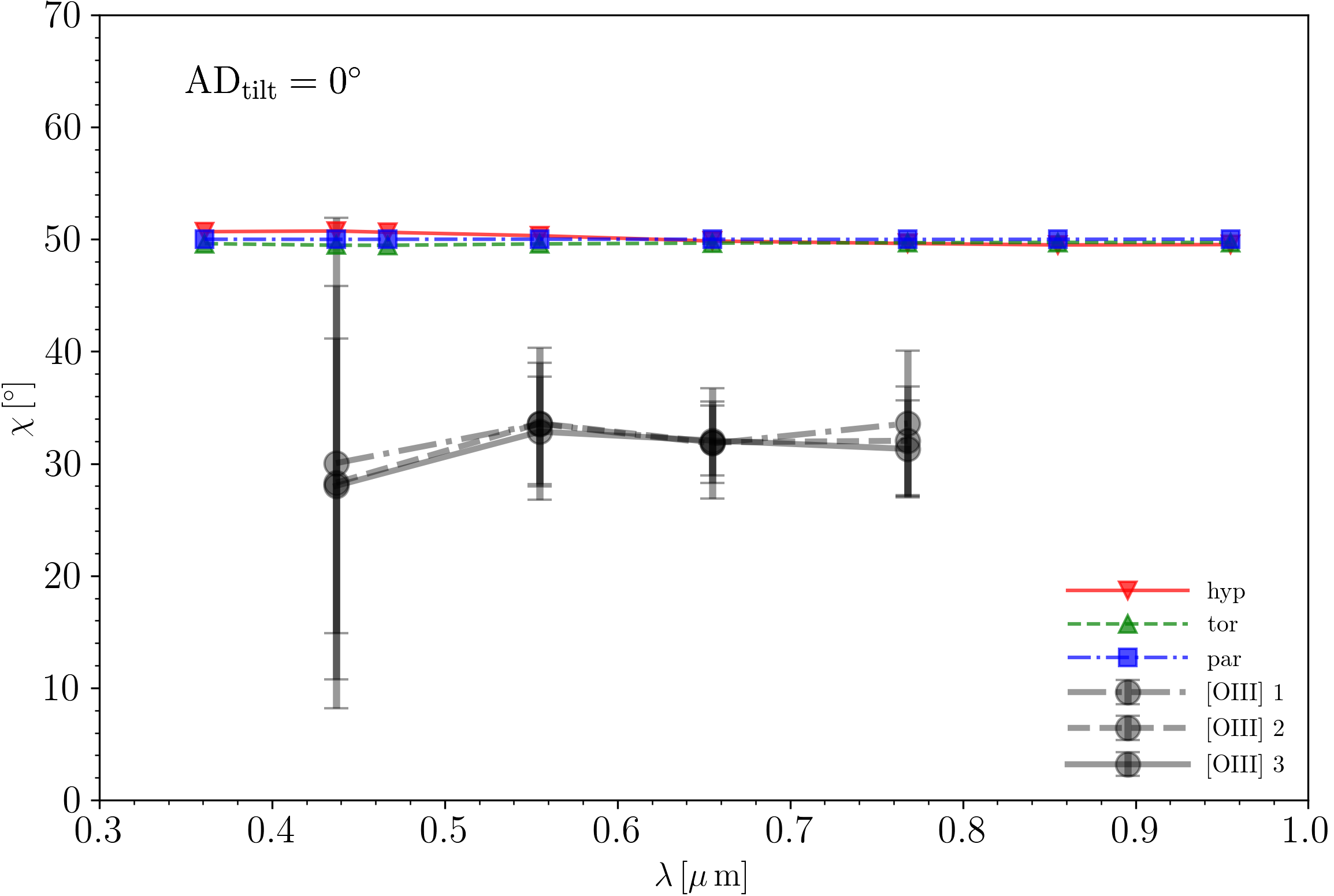}\\
 \includegraphics[width=0.49\textwidth]{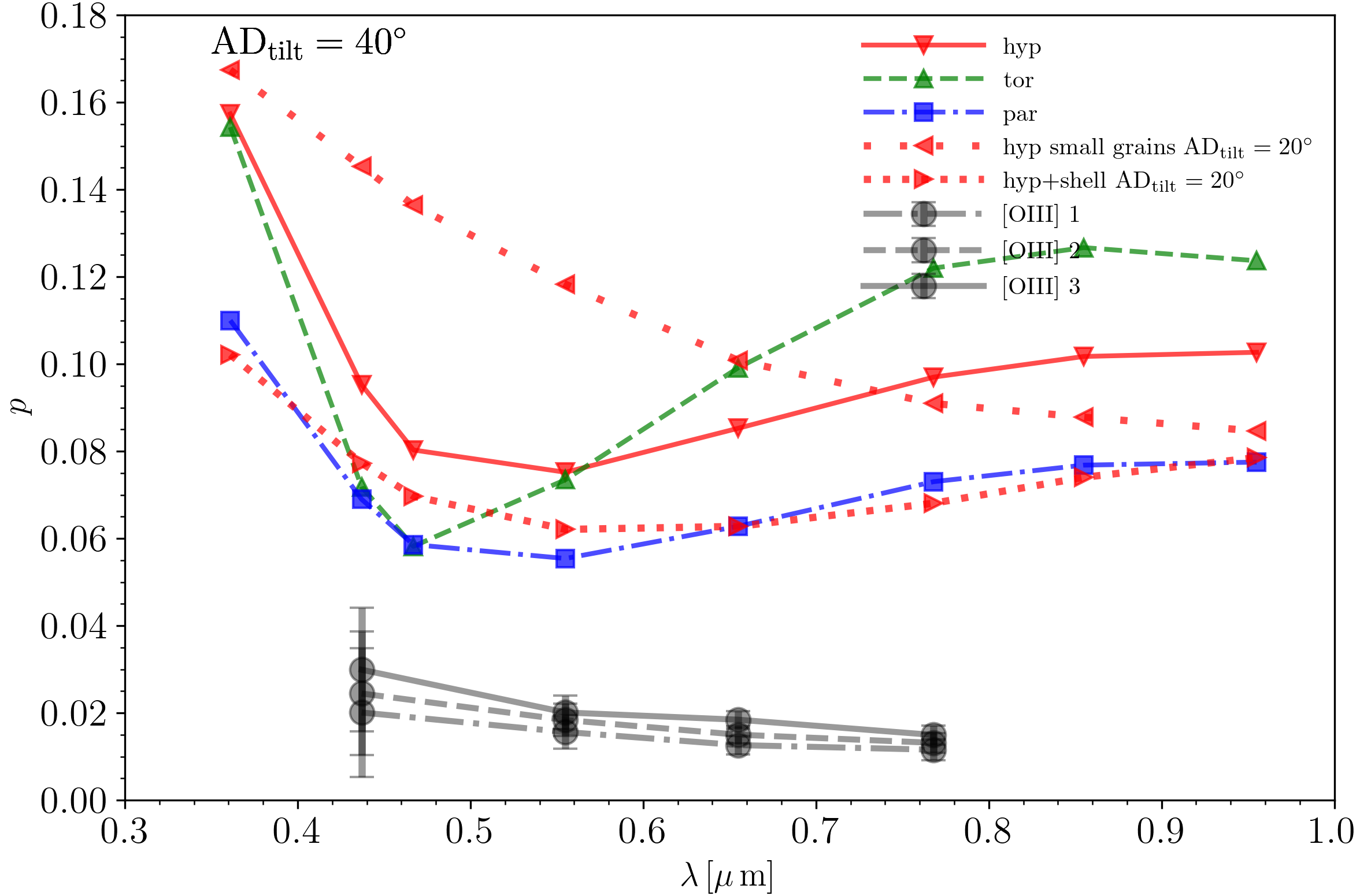}
 \includegraphics[width=0.49\textwidth]{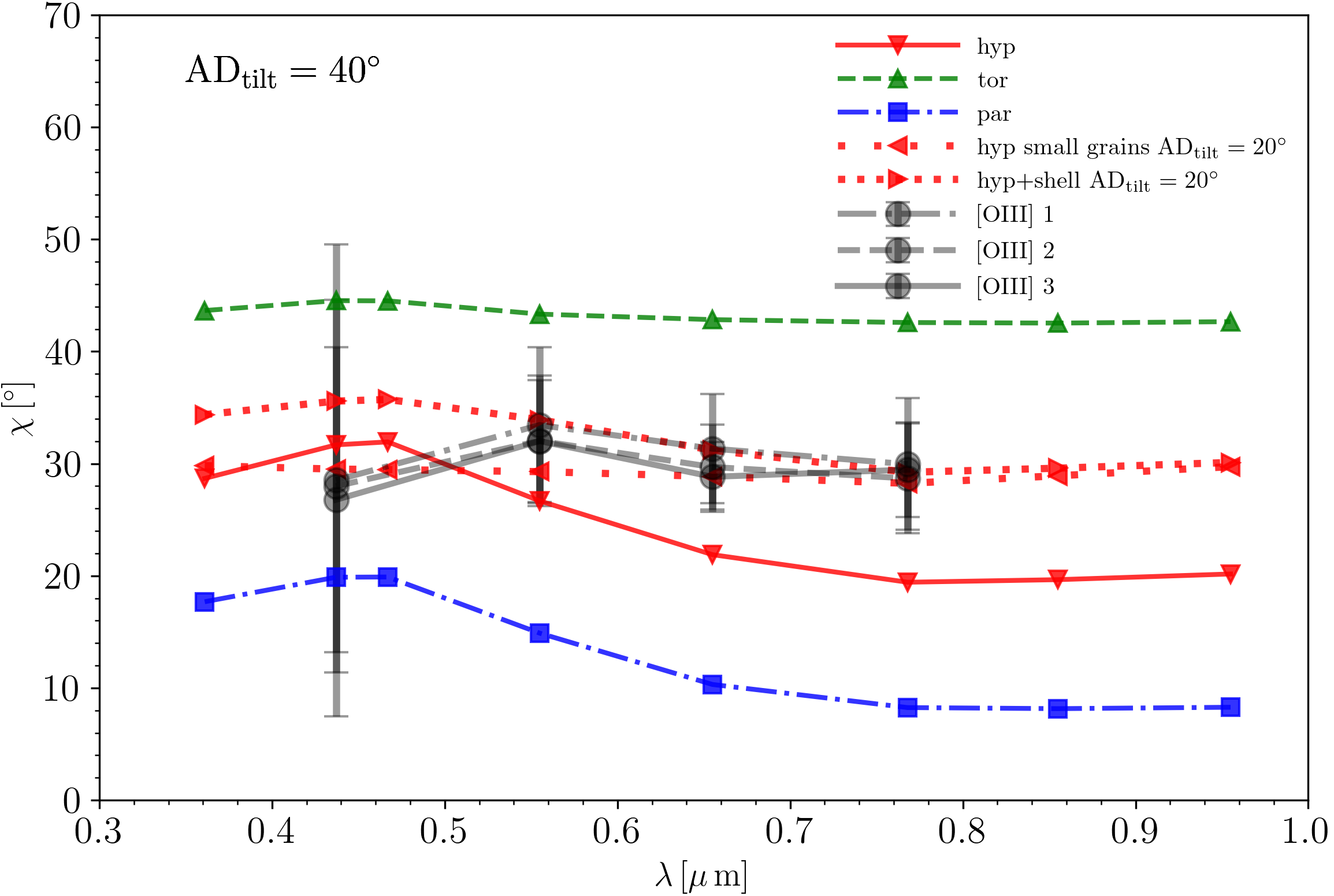}
 \caption{Integrated polarization degree (left) and angle (right) of the models, for the accretion disc aligned with the polar axis (top) and tilted by $40\degr$ (bottom). The three model geometries are shown in red solid (\texttt{disc+hyp}), blue dash-dotted (\texttt{disc+par}) and green dashed (\texttt{tor+sph}) lines. In all panels, black dots are observed values integrated within the three [\ion{O}{iii}] VLT/MUSE contours shown in the Fig.~\ref{fig:obsPolMap} (integration is preformed on Stokes $Q$ and $U$). Two additional red dotted lines represent \texttt{disc+hyp} with \ADtilt{20}. Loosely dotted line with left pointing triangle symbols is a model which included small grains (down to $0.005$ \micron), the densely dotted line with right pointing triangle symbols includes an optically thin sphere shell ($\tau_V=0.1$) enveloping the \texttt{disc+hyp}, representing ambient dust in the host galaxy. Prior to integration, the models have been rotated to match the on-sky orientation of Circinus AGN.}
 \label{fig:obsModInt}
\end{figure*}

The region of our interest, depending on the wavelength, corresponds to only several resolution elements (see the median FWHM of the PSF in each band indicated in Fig.~\ref{fig:obsPolVec} and exact values in table \ref{tab:obs}). Thus, rigorous morphological comparison is not possible. Instead, we opt for comparing integrated values of the data and the models in Fig.~\ref{fig:obsModInt}. The observed Stokes $Q$ and $U$ are integrated within the three levels of the VLT/MUSE [\ion{O}{iii}] contours (see Figs.~\ref{fig:obsPolVec} and~\ref{fig:obsPolMap}). The extent of the models correspond to the smallest and middle contour, but we also consider another, larger region, to examine if and how the polarization properties change farther away from the nucleus. For the models, we tested integration of the full maps and only a single cone above the disc, finding marginal differences. This is due to the almost edge-on view for which the Stokes maps of the models are almost symmetrical with respect to the disc plane. Before integration, the models are rotated to match the on-sky orientation of the Circinus AGN. Aside from the observed $BVRI$ bands, we extended the model range both at shorter and longer wavelengths to serve as a reference for potential future studies. As in the previous section, we test both cases of accretion disc aligned with the cone axis and tilted to its side by $40\degr$.

The integrated polarization degree in the ionization cone ($\approx1.5\%$ in the $I$-band, $\approx3\%$ in the $B$-band within the most compact [\ion{O}{iii}] contour, solid line labelled \#3) is in agreement with low percentages typically found for type 2 sources \citep{Martin1983, Kay1994, Smith2002}. Notably, all the models predict higher values of the polarization degree by at least a factor of few. 
We emphazise that the dilution of polarization due to host starlight can be substantial and it is crucial to remove it to study the true polarization of AGN light \citep{MillerAntonucci1983, Tran1995}. If the fraction of stellar light cannot be accounted for reliably, only a lower limit to the AGN polarization can be inferred \citep{Antonucci2002}.
However, unpolarized stellar light usually dominates the optical continuum emission in type 2 AGN.
Following \citet{Tran1995}, \citet{RamosAlmeida2016} fitted stellar templates to estimate the fraction of starlight in the total observed optical spectrum of Circinus. They found that stellar component accounts for almost all the emission in the $B$ and $V$ bands and concluded that, given the considerable uncertainties in the method \citep[see][and references therein]{Tran1995}, the level of the actual AGN continuum polarization in this galaxy cannot be estimated.

We have made two attempts to estimate the stellar fraction. In the analysis of our VLT/MUSE NMF data for a separate publication (Kakkad et al, subm.), we have used the \textsc{LZIFU} spectral tool \citep{Ho2016} to fit stellar population models at each pixel, which allowed us to infer the ratio of stellar to AGN light (albeit, only $V$ and $R$ bands of FORS2 are within the MUSE NFM range). In addition, we also tried a novel approach by masking the ionization cone area and applying integrated nested Laplace approximation (INLA): an approximate method for Bayesian inference, coupled to Gaussian Markov Random Fields to detect and reconstruct the non-random spatial structure. INLA is a powerful method that can reveal the underlying spatial structure of different types of data \citep{Rue2009, Rue2017}. It has been widely used in various fields, and in astronomy it has been tested on integral field unit observations of galaxies \citep{Gonzalez-Gaitan2019}, demonstrating its capabilities to recover structures otherwise hidden, even with highly sparse spatial information. In this approach, we masked the ionization cone area of the total flux images in each of the four FORS2 bands and allowed INLA to reconstruct the missing part. Since the AGN emission is anisotropic and collimated by the obscuring dust, light outside the ionization cone should be of pure stellar origin. Thus, the reconstructed flux in the masked area provides an estimate of host starlight. Both methods provide very high stellar fractions ($>0.95$), consistent with flux in these bands being fully dominated by the starlight within the uncertainties of each method. Thus, the starlight dilution cannot be removed reliably and we must treat the reported polarization degrees in each band as only lower limits.

\citet{Alexander2000} used an empirical correlation between the $R$-band continuum and broad H$\alpha$ flux for a sample of Seyfert 1 galaxies to constrain the continuum polarization of Circinus AGN; however, they were only able to estimate a lower limit of $9.6\%$.
The most reliable method to determine true continuum polarization due to scattering is the ratio of polarized to total flux across broad emission lines. Since in type 2s the total broad line fluxes are not available, \citet{Marin2014} suggested to compare the typical equivalent width of the polarized broad Balmer emission lines in type 1s \citep[H$\alpha$ EW$\approx 400$ \AA,][]{Smith2002} to the values observed in a given type 2 object. Applying this method to the spectropolarimetry of \citet{Alexander2000}, \citet{Marin2014} inferred a value of $22.4\%$ for Circinus. 
This is similar to the value estimated by \citet{Oliva98} based on a toy model of transmission through the host galaxy disc ($\sim25\%$). However, we stress that these estimates are only a first-order approximation; the actual optical continuum polarization due to scattering in Circinus could be anywhere within the range of $\approx1.6-25\%$ in the $R$-band and $\approx3-35\%$ in the $B$-band.
Thus, starlight dilution in this galaxy prevents us from a quantitative comparison of the observed polarization level and that predicted by the models.

Examining the wavelength dependence in Fig.~\ref{fig:obsModInt}, we see that the polarization degree of the \texttt{tor+sph} is steeply rising with increasing wavelength, in a stark discrepancy to the data. On the other hand, \texttt{disc+hyp} and \texttt{disc+par} models show similar trend to the data, mildly dropping from $B$ to $V$ band and remaining flat towards $R$ and $I$ bands in \ADtilt{0} case. For \ADtilt{40}, both models are rising towards longer wavelength, however, not as steeply as the \texttt{tor+sph} model.

Looking at the polarization angle in the same figure, it is most striking that all the models fail to match the data in the case of the aligned accretion disc. For the tilted disc, the \texttt{disc+hyp} model provides the best match, while the \texttt{disc+par} model predicts lower, and \texttt{tor+sph} higher polarization angle values. The polarization angle of the \texttt{disc+hyp} model drops below the observed values at longer wavelengths.

We do not endeavour an extensive parameter space exploration and fitting, but rather aim for a study focused on qualitative comparison of the three representative model geometries. However, we remind the reader that single dish and interferometric MIR observation strongly favour the \texttt{disc+hyp} model and rule out the alternatives. The fact that the fiducial \texttt{disc+hyp} provides the best match to the optical polarimetry, rounds up a compelling case for this model. Thus, we further examined if small adjustments of the model can match the data even better by: (a) varying the minimal and maximal dust grain sizes, (b) including an additional component -- a spherical shell -- representing host galaxy ISM and (c) changing the accretion disc tilt. We found that including the grains smaller than $0.1$ $\micron$ (down to $0.005$ $\micron$) flattens out the wavelength dependence of the polarization angle. Inclusion of the spherical shell with $\tau_{9.7}=0.1$ has a similar effect. In both cases, the polarization angle is reduced, which can be compensated by the change of accretion disc tilt (\ADtilt{20}). This is the limiting value of the $\text{AD}_{\text{tilt}}$ allowed by the MIR data: lower tilt values would start to reveal the other side of the dusty cone, inconsistent with the bar-like morphology seen by VISIR. Given the uncertainties of the observed values, it is unclear if these additions are necessary: even the starting \texttt{disc+hyp} with reduced accretion disc tilt is within the uncertainties. From this grid of models, we pick two representatives and show them in the bottom panels of the Fig.~\ref{fig:obsModInt} with red loosely and densely dotted lines. Both of these models match the observed polarization angle and provide qualitatively better wavelength dependence of the polarization degree than the other cases. Small grains increase the absolute polarization level, while the model with a shell of ambient dust (\texttt{hyp+shell}) lowers it.
The optical depth of this shell is $\sim20$ times lower than the inferred foreground screen by \citet{Roche2006}. Such a homogeneous optically thick shell of dust would extinguish the unpolarized optical signal from the ionization cone and would not produce consistent MIR morphology; this ambient dust has to be patchy and consist of a large number of small clumps to allow the transmission. \textsc{skirt} code is primarily designed to calculate self-consistently absorption and thermal dust emission, which requires each clump to be sufficiently resolved by the numerical grid of cells containing dust. In this particular case, resolving a large number of small clumps requires large amount of memory, making this a challenging task for our current computational resources. However, \citet{Marin2015} explored fragmentation of the polar scattering medium and found that for higher filling factors the polarization is increased, while for low values, it remains consistent with homogeneous case. Thus, it is a reasonable assumption that polarization signature of such an clumpy ambient dust would not differ significantly from our case of homogeneous \texttt{hyp+shell} model. Additionally, we found that polarization degree can be lowered further, and almost match the observed level of $\approx1-3\%$, in the cases of models with smaller inclination ($i\approx80\degr$) or reduced angular width of the hyperboloid shell ($\approx3\degr$). However, given the large uncertainties due to the starlight dilution, we refrain from further adjustments of the model, since even the original  \texttt{disc+hyp} model is consistent with estimated intrinsic level of polarization ($\approx9-25\%$ \citealt{Oliva98,Alexander2000,Marin2014}).

We note that in all our models we considered only scattering on dust grains, neglecting electrons. This is justified by the observed change of polarization with wavelength in the region of our interest (within and around the ionization cone), as the scattering on electrons results in wavelength-independent polarization. Exploring qualitatively similar model setup in MCRT approach, \citep{Wolf-Henning1999} found that wavelength dependence can be produced by optically thin cones, while the presence of electrons only increases level of polarization.

We have also constructed full 3D models of the Circinus galaxy: a simple model based on the analytical description and a realistic one. The former is a three-component model, consisting of a double-exponential disc to describe the stellar and dust discs, and a central bulge with a Sersic profile, with parameters corresponding to the typical local galaxies \citep{DeGeyter2014}. The realistic model is based on the Herschel/SPIRE $250$ $\micron$ map which is deprojected and added an exponential vertical profile. The results of radiative transfer simulations of these models shall be presented in the second publication which will be focused on the properties of the polarization and dust of the host galaxy. Since this is an on-going work, here we just note that, for the central region of our interest, the preliminary results are qualitatively consistent with our simplified host ambient dust in a form of a sphere shell surrounding the AGN model described above.

We did not extend our radiative transfer simulations to the NIR, since \textsc{skirt} code does not yet support dichroic absorption due to non-spherical dust grains aligned in the magnetic and/or radiation field, potentially dominant polarization mechanism in this wavelength domain. However, it is useful to briefly consider the results of \citet{Ruiz2000} who obtained $JHK$ bands polarimetric maps of Circinus with Anglo-Australian Telescope. Based on their analysis, they argue that scattering is responsible for polarization in the $J$-band, while dichroism becomes dominant in the $K$-band (although $J$-band as well might have contribution from dichroism at larger distances). The polarization position angle we find with FORS2 in the studied region ($31.3\pm4.3$ in the $I$-band) is consistent with the value reported by \citet{Ruiz2000} in their $J$-band $1\arcsec$ aperture ($34.1\pm3.8$).

Having in mind everything laid out in this section, we conclude that the \texttt{disc+hyp} model based on the MIR data is consistent with the optical polarimetry observations obtained with VLT/FORS2 imaging.
Namely, the polarization angle makes a clear distinction between the different model geometries and orientations of the primary source, prefering the \texttt{disc+hyp} illuminated by a tilted accretion disc (Fig.~\ref{fig:obsModInt}). Due to the dilution by the unpolarized starlight, the observed polarization degree can be considered only as a lower limit. However, detection of broad lines in the polarized spectrum of Circinus \citep{RamosAlmeida2016} is an evidence of scattering being the main mechanism producing the polarization in the region under the study here. Given that MIR observations revealed presence of dust in the polar region, it is plausible to assume that the observed wavelength dependence of the polarization degree in the $BVRI$ bands is due to the scattering on dust grains as well. Thus, even though quantitative comparison is hindered, qualitative analysis presented above deserves credibility.

We remind the reader that hyperboloid shape of the wind is pronounced at the scale of a few parsecs; beyond that, the hyperboloid effectively follows the shape of a cone. At the achieved angular resolution, FORS2 observations cannot probe the parsec scale structure. Thus, if we would base the conclusions only on the here presented analysis of polarimetric data, the best model should be described as \texttt{disc+cone}. However, as already noted in Sec.~\ref{sec:intro} and \ref{sec:mod}, interferometic data and numerical simulations provide strong evidences for the hyperboloid shape of the wind on small scales \citep{Stalevski2019,Venanzi2020,Isbell2022}.

\section{Summary and Conclusions}
\label{sec:sum}

We presented new $BVRI$ imaging polarimetry of the nucleus and ionization cone in the Circinus galaxy with the highest angular resolution up to date: 4.89 pc/pix, with PSF FWHM in the range of $0.68\arcsec$ to $1.34\arcsec$, depending on the wavelength (at the distance of 4.2 Mpc, $1\arcsec\approx20$ pc).  Circinus is the nearest Seyfert 2 galaxy harbouring an archetypal obscured AGN. Recent MIR interferometry \citep{Tristram2014} and single dish imaging \citep{Asmus2016}, together with detailed radiative transfer modelling \citep{Stalevski2017,Stalevski2019}, have cast this galaxy in a major role as a prototype in the emerging population of `polar dust AGN', in which a major fraction of the MIR emission is associated with dusty winds blown away from the sublimation zone by the radiation pressure \citep{Honig2019}. In this third paper in the series, we dissected the Circinus AGN using the VLT/FORS2 polarimeter in the imaging mode. We performed a limited parameter study of the polarization signatures produced by three different dust geometries. Namely, a model based on our previous works, consisting of a compact thin disc with a hyperboloid shell (effectively a hollow cone at here considered scales) (\texttt{disc+hyp}), and two plausible alternatives: a disc with a paraboloid shell (\texttt{disc+par}), and a torus with a sphere shell (\texttt{tor+par}).
The former model has a strong hold in radiative transfer modeling \citepalias{Stalevski2019}, numerical simulations of dusty gas based of a semi-analytical model \citep{Venanzi2020} and MIR interferometric imaging \citep{Isbell2022}. The latter two alternatives were included to explore if, in principle, polarimetric observations alone could distinguish between the different geometries.

Our focus is on the AGN core and ionization cone, up to 40 pc ($\approx2\arcsec$) from the nucleus. More precisely, comparison of the models and the data is limited to the region corresponding to the small and middle [\ion{O}{iii}] MUSE contours (Kakkad et al, subm.) overplotted in the observed polarization maps. From our analysis and comparison of the data to the radiative transfer models, we highlight the following:

\begin{enumerate}

    \item Maps of the polarization degree, position angle and Stokes parameters reveal a (bi-)conical morphology, consistent with the radiation from the central obscured source being scattered on the dust within the ionization cone of AGN.

    \item The polarization degree is wavelength-dependent, indicating that scattering on dust grains is the main mechanism of polarization. This is supported by the presence of broad lines in the polarized spectrum \citep{RamosAlmeida2016}.
    
    \item The polarization degree (from $\approx3\%$ in the $B$-band to $\approx1\%$ in the $I$-band, Table~\ref{tab:con}) is lower than predicted by the models. However, large uncertainties prevent quantitative comparison. Namely, unpolarized stellar emission dominates the total flux in the optical range. Thus, the observed polarization level can be considered only a lower limit. We were unable to robustly account for the dilution by starlight, but if approximate estimates from the literature are applied, the actual polarization produced by scattering of AGN light is consistent with the model expectations, albeit, within considerable uncertainty.
    
    \item The polarization angle ($\approx30\degr$, Table~\ref{tab:con})) is reproduced by the \texttt{disc+hyp} model with the anisotropic accretion disc tilted by $20-40\degr$ towards the edge of the cone. An accretion disc aligned with the cone axis, and alternative model geometries (\texttt{disc+par} and \texttt{tor+sph}) are inconsistent with the data.
    
\end{enumerate}

We conclude that the model of a thin disc and a hyperboloid shell illuminated by a tilted accretion disc represents well the dust structure in the active nucleus of the Circinus galaxy. 
To be precise, the hyperboloid effectively resembles a hollow cone beyond a few parsecs. At here achieved angular resolution, FORS2 observations cannot distinguish between these two geometries; hyperboloid shape is established primarily by the MIR interferometry coupled with radiative transfer modeling.
Observations in a range of wavelengths and techniques -- from X-ray and optical, to MIR and maser emission -- paint a consistent picture, which is now corroborated by the evidence from optical polarimetric imaging.

\section*{Acknowledgements}

M.S. and \DJ.S. acknowledge support by the Science Fund of the Republic of Serbia, PROMIS 6060916, BOWIE and by the Ministry of Education, Science and Technological Development of the Republic of Serbia through the contract No.~451-03-9/2022-14/200002. S.G.G and A.M. acknowledge support from FCT under Project CRISP PTDC/FIS-AST-31546/2017 and Project~No.~UIDB/00099/2020.
DA acknowledges funding through the European Union’s Horizon 2020 and Innovation programme under the Marie Sklodowska-Curie grant agreement no. 793499 (DUSTDEVILS). \DJ.S. acknowledges supported by the F.R.S.
FNRS under grant PDR T.0116.21. The work was also supported by the programme of scientific and technological cooperation between the government of the Republic of Serbia and the government of the Republic of Portugal, (grant~No.~337-00-00227/2019-09/53 and FCT 5581 DRI, Sérvia 2020/21). Some of the computations were performed at the cluster “Baltasar-Sete-Sóis” and supported by the H2020 ERC Consolidator Grant "Matter and strong field gravity: New frontiers in Einstein’s theory" grant agreement no. MaGRaTh-646597.


\section*{Data Availability}

Based on European Southern Observatory (ESO) observing programmes 0101.B-0647(A) and 0103.B-0517(A).




\bibliographystyle{mnras}
\bibliography{references} 



\appendix

\section{Calculation of Stokes parameters and uncertainties}
\label{app:stokes}

A given field observed in the four half-wave plate (HWP) positions at constant intervals of $\Delta \theta=22.5\deg$ results in 4 images of ordinary ($f_O$) and 4 images of extraordinary ($f_E$) beams, as described in the Sec.~\ref{sec:reduction}. 
The double difference method used in this work to calculate the polarization degree and polarization angle relies on normalised flux differences $F_a(x,y)$ evaluated at the four HWP position angles $a=0^\circ ; 22.5^\circ; 45^\circ ; 67.5^\circ$ and as a function of pixel position $x,y$ \citep[e.g.][]{Patat06}:

\begin{eqnarray}
\label{eq:FNi}
F_a(x,y) \equiv \frac{f_{O,a} - f_{E,a}}{f_{O,a} + f_{E,a}} \, \,  . 
\end{eqnarray}

\noindent The intensity at each position in the sky is estimated using the average of the sum of the ordinary and extraordinary beams in different observations,  taken in the corresponding pixel position $x,y$: 
\begin{eqnarray}
I(x,y) =  \langle f_{O,a}(x,y) + f_{E,a}(x,y)\rangle \, \,  . 
\end{eqnarray}

This enables the calculation of the normalised Stokes parameters: 

\begin{eqnarray}\label{eq:qu}
q(x,y)  &=& \frac{1}{2} \; \left(F_{0} - F_{45.0}\right) \\
u(x,y)  &=& \frac{1}{2} \; \left(F_{22.5} - F_{67.5}\right).
\end{eqnarray}

From this, the polarization degree and polarization angle are given by:

\begin{eqnarray}\label{eq:pol}
p(x,y) &=& \sqrt{q^2 + u^2} \\
\chi(x,y) &=& \frac{1}{2} \arctan \frac{u}{q} \, \, .
\end{eqnarray}

The uncertainties at every pixel of the ordinary and extraordinary beams come from the quadrature sum of the sky noise (which comes from the variance of a source-masked sigma-clipped average of the sky) and the Poisson error:

\begin{eqnarray}
\sigma_f(x,y) = \sqrt{\sigma_{\rm bkg}^2+f/g}\, \, ,
\end{eqnarray}
where $f$ is the ordinary or extraordinary flux at a given HWP angle and $g$ is the electron-to-count gain. This uncertainty per pixel is used both, when doing photometry for field stars, and for individual pixel analysis. When several images of the same field, filter and HWP angle are present, we take the median and median absolute deviation of all of them as the new flux and associated error, respectively.

The flux uncertainties are then propagated through to the Stokes parameters, as follows:

\begin{eqnarray}
    \sigma_q(x,y) &=& \frac{1}{2}\sqrt{\sigma_{F_0}^2+\sigma_{F_{45.0}}^2}\\ \sigma_u(x,y) &=& \frac{1}{2}\sqrt{\sigma_{F_{22.5}}^2+\sigma_{F_{67.5}}^2}\, \, ,
\end{eqnarray}
where:
\begin{eqnarray}
    \sigma_{F_a}(x,y) = \sqrt{\frac{(2f_{E,a}\sigma_{f_{O,a}})^2 + (2f_{O,a}\sigma_{f_{E,a}})^2}{(f_{O,a}+f_{E,a})^2}}\, \, .
\end{eqnarray}

For multiple $q$ and $u$ values obtained from different offsets, the final values and uncertainties are the median, $q(x,y) = <q_i>$, and median absolute deviation of all offsets $i$. The observed $q$ and $u$ values are corrected for various instrumental and foreground effects with Eq.~\ref{eq:stokescorr} propagating the corresponding errors.

Finally, the uncertainties in $p$ and $\chi$ are propagated as:

\begin{eqnarray}
    \sigma_{p}(x,y) &=& \sqrt{\frac{(q\sigma_q)^2 + (u\sigma_u)^2}{q^2+u^2}}\\
    \sigma_{\chi}(x,y) &=& \frac{1}{2}\sqrt{\frac{(q\sigma_u)^2 + (u\sigma_q)^2}{q^2+u^2}}\, \, .
\end{eqnarray}



\bsp	
\label{lastpage}
\end{document}